\documentclass[preprint, amsmath, amssymb, aps]{revtex4-2}%
\usepackage{adjustbox}
\usepackage{amssymb}
\usepackage{amsfonts}
\usepackage{amsmath}
\usepackage{graphicx}%
\usepackage{dcolumn}
\usepackage[utf8]{inputenc}
\usepackage[T1]{fontenc}
\newtheorem{theorem}{Theorem}

\newtheorem{lemma}[theorem]{Lemma}

\newenvironment{proof}[1][Proof]{\noindent\textbf{#1.} }{\ \rule{0.5em}{0.5em}}
\ifx\pdfoutput\relax\let\pdfoutput=\undefined\fi
\newcount\msipdfoutput
\ifx\pdfoutput\undefined\else
\ifcase\pdfoutput\else
\ifx\paperwidth\undefined\else
\ifdim\paperheight=0pt\relax\else\pdfpageheight\paperheight\fi
\ifdim\paperwidth=0pt\relax\else\pdfpagewidth\paperwidth\fi
\fi\fi\fi
\begin{document}
\title{Generalized Lie Symmetries and Almost Regular Lagrangians: A Link
  Between Symmetry and Dynamics} 
\author{Achilles D. Speliotopoulos}
\affiliation{Department of Physics, University of California, Berkeley, CA 94720 USA}
\altaffiliation[Also at ]{Physical Science and Engineering Division,
  Diablo Valley College, Pleasant Hill, CA 94523, USA} 
\email{ads@berkeley.edu}
\date{\today}

\nopagebreak

\begin{abstract}
  The generalized Lie symmetries of almost regular Lagrangians are
  studied, and their impact on the evolution of dynamical systems
  is determined. It is found that if the action has a generalized Lie
symmetry, then the Lagrangian is necessarily singular; the converse is
not true, as we show with a specific example. It is also found that
the generalized Lie symmetry of the action is a 
Lie subgroup of the generalized Lie symmetry of the Euler-Lagrange
equations of motion. The converse is once again not true, and there are systems
for which the Euler-Lagrange equations of motion have a generalized Lie
symmetry while the action does not, as we once again show through a specific
example. Most importantly, it is shown that each
generalized Lie symmetry of the action contributes one arbitrary
function to the evolution of the dynamical system. The number of such symmetries
gives a lower bound to the dimensionality of the family of curves
emanating from any set of allowed initial data in the Lagrangian phase
space. Moreover, if second- or higher-order Lagrangian constraints are
introduced during the application of the Lagrangian constraint algorithm,
these additional constraints could not have been due to the
generalized Lie symmetry of the action.     
 
\end{abstract}

\maketitle

\section{Introduction\label{&Intro}}

The symmetries of the Euler-Lagrange equations of motion
were recently used to study the constrained dynamics of singular
Lagrangians \cite{ADS2020}. The focus was on almost regular
Lagrangians \cite{Got1978, Got1979, Got1980, Car1990a}, and it was
found that for these Lagrangians the Euler-Lagrange equations of motion  
admit a generalized Lie symmetry (also known as a local gauge
symmetry). The generators $\mathcal{S}\hbox{ym}$ of this symmetry group
$\hbox{Gr}_{\mathcal{S}\hbox{ym}}$ were determined in the Lagrangian
phase space approach to Lagrangian mechanics, and were found to lie
in the kernel of the Lagrangian two-form $\mathbf{\Omega}_L$. While it
is well-known that the solutions $\mathbf{X}_E$ of the energy equation,  
\begin{equation}
  0=\mathbf{d}E-i_{\mathbf{X}_E}\mathbf{\Omega}_L,
  \label{EnergyE}
\end{equation}
is not unique for almost regular Lagrangians, it was shown in \cite{ADS2020} that
the action of $\mathcal{S}\hbox{ym}$ on a general solution  
to this equation\textemdash and in particular, on the \textbf{second-order,
  Lagrangian vector field} (SOLVF)\textemdash will result in a vector
field that is no longer a solution of Eq.~$(\ref{EnergyE})$. Thus, not all
solutions of the energy equation have
$\hbox{Gr}_{\mathcal{S}\hbox{ym}}$ as a symmetry group. It is,
however, possible to construct 
solutions to Eq.~$(\ref{EnergyE})$ for whom $\mathcal{S}\hbox{ym}$
does generate a group of symmetry transformations
\cite{ADS2020}. These vector fields are called \textbf{second-order,
  Euler-Lagrange vector fields} (SOELVFs). As the 
evolution of the dynamical system for singular Lagrangians must lie on
Lagrangian constraint surfaces \cite{Car1990a}, a Lagrangian constraint
algorithm for SOELVFs was also introduced in \cite{ADS2020} to
construct such solutions to the energy equation. It was
then shown that these SOELVFs, along with the dynamical structures
in the Lagrangian phase space needed 
to describe and determine the motion of the dynamical system, are projectable to the
Hamiltonian phase space. In particular, the primary
Hamiltonian constraints can be constructed from vectors that lie in the kernel of
$\mathbf{\Omega}_L$, and the Lagrangian constraint
algorithm for the SOELVF is equivalent to the stability analysis of 
the total Hamiltonian (we follow the terminology found in
\cite{Hen1992}; see also \cite{Dir1950, Mun1989, Lus2018}) obtained
using constrained Hamiltonian mechanics. Importantly, the end result
of this stability analysis gives a Hamiltonian vector field that is the 
projection of the SOELVF obtained from the Lagrangian constraint
algorithm. The Lagrangian and Hamiltonian formulations of mechanics
for almost regular Lagrangians were thereby shown to be equivalent. 

While \cite{ADS2020} focused on the generalized Lie symmetries of the
Euler-Lagrange equations of motion and whether the dynamical
structures constructed in the Lagrangian phase space are projectable to
the Hamiltonian phase space, in this paper the focus is on the symmetries
of the action itself and the impact these symmetries have on the
evolution of dynamical systems. This impact is found to be quite broad,
surprisingly restrictive, and unexpectedly subtle. Indeed, even the
seemingly reasonable expectation that any generalized Lie symmetry of
the Euler-Lagrange equations of motion should be a reflection of the
symmetries of the action itself is not borne out.

We find that if the action has a generalized Lie
symmetry, then its Lagrangian is necessarily singular; the converse
need not be true, as we show through a specific example. We also find
that the generators of the generalized Lie symmetry of the action form
a Lie sub-algebra of the generators of the
generalized Lie symmetry of the Euler-Lagrange equation of motion;
once again, the converse is not true. We give an example of a dynamical
system for which the Euler-Lagrange equations of motion has a
generalized Lie symmetry, while its action does not. Most importantly,
for systems where the Lagrangian is almost regular and for 
which the two-form $\mathbf{\Omega}_L$ has constant rank, we show that
each generalized Lie symmetry of the action contributes one
arbitrary constant to the SOELVF. The dimensionality of the space of
solutions to the energy equation that have
$\hbox{Gr}_{\mathcal{S}\hbox{ym}}$ as a symmetry group is thus at
least as large as the number of generalized Lie symmetries of the
action. Moreover, if second- or higher-order Lagrangian
constraints are introduced during the application of the Lagrangian
constraint algorithm, these additional constraints cannot be due to the
generalized Lie symmetry of the action.      
      
Symmetries of Lagrangian systems have been studied before. However,
such analyses have been focused on time-dependent Lagrangians \cite{Pri1983, Pri1985, 
Cra1983, Car1991, Car1988b, Car1992, Car1993, Car2003}; on systems of
first-order evolution equations \cite{Car1990b, Mar1992, Gra2002, Gra2005,
  Pop2017}; or on general solutions of Eq.~$(\ref{EnergyE})$ 
\cite{deL1995} (see also \cite{Dim2016}). Importantly, the great majority of these
studies have been done using first-order prolongations on
first-order jet bundles with a focus on the Lie symmetries of first-order
evolution equations. Our interest is in the symmetries of the action, which
naturally leads us to consider generalized Lie symmetries and
second-order prolongations. To our knowledge, such symmetry analysis
of the action has not been done before. (The framework for
$k^{\hbox{\textit{th}}}$-order prolongations on
$k^{\hbox{\textit{th}}}$-order jet bundles have been introduced before
\cite{Car1993, deL1995, Car2003, Pop2009, Pop2011}, but they were not applied
to the action or to the Euler-Lagrange equations of motion.)

The rest of the paper is arranged as follows. In \textbf{Section
  \ref{&Sym}} the conditions under which the action for a dynamical
system, and the conditions under which the Euler-Lagrange equations of
motion for this action, have a generalized Lie symmetry are
determined. To compare the conditions for each, the analysis for the
two are done separately, with each self-contained. In
\textbf{Section \ref{&review}} properties of the Lagrangian phase
space are reviewed, and the notation used here established. The
generators of the generalized Lie symmetry group for the
Euler-Lagrange equations of motion were determined in \cite{ADS2020},
and a summary of the results found therein that are needed here is
given. In \textbf{Section \ref{&A-S}} the generators of the
generalized Lie symmetry group for the action is found within the
Lagrangian phase space approach, and their relation to the
generators for the symmetry group of the Euler-Lagrange equations of
motion is determined. The impact of the symmetries of the action on
the SOELVF is then analyzed by applying the 
Lagrangian constraint algorithm introduced in \cite{ADS2020} to these
SOELVF. The results obtained in this paper is then applied to three different
dynamical systems in \textbf{Section \ref{&Exam}}. In particular, an
example of a dynamical system that has no generalized Lie symmetries
and yet is still singular, and another example where the action has no
symmetries and yet the Euler-Lagrange equations of motion do, are
given. Concluding remarks can be found in \textbf{Section \ref{&Conc}}.      

\section{Generalized Lie symmetries and Lagrangian Mechanics\label{&Sym}}

In this section we determine the conditions under which the action of
a dynamical system, and the conditions under which the Euler-Lagrange
equations of motion for this system, has a generalized Lie
symmetry. While the determination for both is done within Lagrangian
mechanics, the analysis for the action is completed separately from that of
the equations of motion\textemdash with each self-contained\textemdash so 
that the two conditions can be compared. We will later show that every
generator of the generalized Lie symmetry of the action is a
generator of a generalized Lie symmetry of the Euler-Lagrange
equations of motion. Interestingly, the converse is not true. 

\subsection{Symmetries of the Action \label{&A-Sym}}

We begin with Lagrangian mechanics, and an analysis of the
generalized Lie symmetry \cite{Olv1993} of the action 
\begin{equation}
  S := \int_{t_1}^{t_2}L\left(q(t),\dot{q}(t)\right)dt,
  \nonumber
\end{equation}
for a dynamical
system on a $D$-dimensional configuration space $\mathbb{Q}$. Here,
$L\left(q(t), \dot{q}(t)\right)$ is the Lagrangian along a path $q(t)
= \left(q^1(t), \dots, q^D(t)\right)$ on $\mathbb{Q}$ with end
points given by $Q_1:=q(t_1), Q_2:= q(t_2)$. These points are 
chosen at the same time the choice of $S$ is made, and are fixed.

As $L\left(q(t),\dot{q}(t)\right)$ depends on both the position $q(t)$
and the velocity $\dot{q}(t)$ of the path, we consider a generalized
Lie symmetry that is generated by 
\begin{equation}
  \mathbf{g}_L := \rho_L(q, \dot{q})\cdot \frac{\mathbf{\partial}
    \>\>\,}{\mathbf{\partial} q}, 
  \nonumber
\end{equation}
where $\rho_L(q,\dot{q})$ does not depend explicitly on
time. Evolution along the path gives the total time derivative 
\begin{equation}
  \frac{\mathbf{d}\>\>\>}{\mathbf{d}t} := \dot{q}\cdot
  \frac{\mathbf{\partial}\>\>\>}{\mathbf{\partial} q} +\ddot{q}\cdot
  \frac{\mathbf{\partial} \>\>\>}{\mathbf{\partial} \dot{q}}.
  \label{Dt}
\end{equation}
This in turn gives $\dot{\rho}_L:=\mathbf{d}\rho_L/\mathbf{d}t$, and the second-order
prolongation vector \cite{Olv1993}, 
\begin{equation}
  \hbox{\textbf{pr }}\mathbf{g}_L := \rho_L \cdot
  \frac{\mathbf{\partial}\>\>\>}{\mathbf{\partial} q} 
  + \dot{\rho}_L \cdot \frac{\mathbf{\partial}\>\>\>}{\mathbf{\partial} \dot{q}} +
  \ddot{\rho}_L\cdot \frac{\mathbf{\partial} \>\>\>}{\mathbf{\partial}
    \ddot{q}},
  \label{prog}
\end{equation}
on the second-order jet space $\mathbb{M}^{(2)}=\{(t, q,
\dot{q},\ddot{q})\}$ where this $\hbox{\textbf{pr
}}\mathbf{g}_L\in\mathbf{T}\mathbb{M}^{(2)}$.

Under this generalized Lie symmetry, the action varies by
\begin{equation}
  \delta S=\int_{t_1}^{t_2}\hbox{\textbf{pr }}\mathbf{g}_L \Big[L(q(t),
  \dot{q}(t))\Big]dt,
  \nonumber
\end{equation}
with the requirement that 
$\rho_L(q(t_1), \dot{q}(t_1))=0=\rho_L(q(t_2),
\dot{q}(t_2))$. Then after an integration by parts,
\begin{equation}
  \delta S=\int_{t_1}^{t_2} \rho_L\cdot\left[\frac{\partial L}{\partial
    q} - \frac{d\>\>\>}{dt}\left(\frac{\partial L}{\partial
    \dot{q}}\right)\right]dt.
  \label{Ae1}
\end{equation}
It is important to realize that the action may be evaluated along any path on
$\mathbb{Q}$. As such, if $\mathbf{g}_L$ generates a symmetry of the action, then
Eq.~$(\ref{Ae1})$ must vanish for \textit{all} paths $q(t)$ on
$\mathbb{Q}$, and not just for those that minimize the action.

To make connection with the Lagrangian phase space approach used in
the rest of the paper, we make use of
\begin{equation}
E\left(  q,\dot{q}\right)  :=\dot{q}^{a}\frac{\partial L\left(  q,\dot
  {q}\right)  }{\partial\dot{q}^{a}}-L\left(  q,\dot{q}\right)  ,
\nonumber
\end{equation}%
along with
\begin{equation}
M_{ab}\left(  q,\dot{q}\right)  :=\frac{\partial^{2}L\left(  q,\dot{q}\right)
}{\partial\dot{q}^{a}\partial\dot{q}^{b}}, \quad \hbox{and }\quad 
F_{ab}\left(  q,\dot{q}\right)  := \frac{\partial^{2}L\left(
  q,\dot{q}\right)  }{\partial\dot{q}^{a}\partial q^{b}} -
\frac{\partial^{2}L\left(  q,\dot{q}\right)}{\partial\dot{q}^{b}\partial 
  q^{a}},
\nonumber
\end{equation}
to express Eq.~$(\ref{Ae1})$ as
\begin{equation}
  \delta S=-\int_{t_1}^{t_2} \rho^a_L\left(\frac{\partial E}{\partial
    q^a} +F_{ab}(q,
  \dot{q})\dot{q}^b+M_{ab}(q,\dot{q})\ddot{q}^b\right)dt.
  \label{e2}
\end{equation}
Here, Latin indices run from $1$ to $D$, and Einstein's summation
convention is used. We then arrive at our first result.

\begin{lemma} \label{Action-Sym} An action $S$ of a dynamical
  system has a generalized Lie symmetry generated by $\mathbf{g}_L$ if
  and only if there exists a $\rho_L\in \hbox{ker }M_{ab}$ such that 
  \begin{equation}
    0=\rho_L^a(q,\dot{q})\left(\frac{\partial E}{\partial
      q^a}+F_{ab}(q,\dot{q})\dot{q}^b\right),
    \label{e3}
  \end{equation}
  on $\mathbf{T}\mathbb{Q}$.
  
  \begin{proof}
    If $\mathbf{g}_L$ generates a generalized Lie symmetry of $S$, then
    Eq. $(\ref{e2})$ must vanish for all paths on $\mathbb{Q}$. For an
    arbitrary path on $\mathbb{Q}$ the
    curvature of the path $\ddot{q}$ will not depend on
    either the $q(t)$ or the $\dot{q}(t)$ for the path, however. As such, 
    for $\delta S=0$, it must be that $\rho^a_LM_{ab}\ddot{q}^b=0$ for any choice of
    $\ddot{q}$, and thus $\rho_L^a\in\hbox{ker }M_{ab}$. The remaining
    terms in Eq.~$(\ref{e2})$ gives the condition
    Eq.~$(\ref{e3})$. 
  \end{proof}
\end{lemma}

The set of all vector fields $\mathbf{g}_L$ that satisfy \textbf{Lemma
  $\mathbf{\ref{Action-Sym}}$} is denoted by  $\mathfrak{g}_L$, while 
$\hbox{\textbf{pr }}\mathfrak{g}_L := \{\hbox{\textbf{pr
}}\mathbf{g}_L\ \vert \ \ \mathbf{g}_L\in \mathfrak{g}_L\}$ is the set of
their prolongations. This $\hbox{\textbf{pr }}\mathfrak{g}_L$ is involutive
\cite{Olv1993}, and the conditions under which $\hbox{\textbf{pr }}\mathfrak{g}_L$
generates a generalized Lie symmetry group are given in
\cite{Olv1993}.

We see from \textbf{Lemma \ref{Action-Sym}} that if the
action has a generalized Lie symmetry, then the Lagrangian is
necessarily singular, and as such the Lagrangian two-form
$\mathbf{\Omega}_L$ will not have maximum rank. It is also important
to note that while equations of the form
Eq.~$(\ref{e3})$ often appear in the Lagrangian phase space description of 
mechanics \cite{ADS2020}, they appear as Lagrangian
constraints, conditions that must be imposed for evolution under the Euler-Lagrange
equations to be well defined. Here, Eq.~$(\ref{e3})$ is not a 
constraint. Rather, because the action must have this symmetry for
\textit{all} possible paths on $\mathbb{Q}$, and since the set of all possible 
paths cover $\mathbb{Q}$, Eq.~$(\ref{e3})$ is a condition on $\rho_L$
that must be satisfied identically on \textit{all} of
$\mathbf{T}\mathbb{Q}$\textemdash and 
thus, on the Lagrangian phase space\textemdash for $\mathbf{g}_L$ to
be a generator of the symmetry group. We will see that not all the vectors in
$\hbox{ker } M_{ab}$ satisfy the identity Eq.~$(\ref{e3})$,
however, and thus not all of these vectors will generate a generalized Lie
symmetry of the action.   
  
\subsection{Symmetries of the Euler-Lagrange
  Equations of Motion\label{&EL-Sym}}

While in \textbf{Section \ref{&A-Sym}} the focus was on arbitrary paths on the
configuration space $\mathbb{Q}$ and the symmetries of the action,
in this section the focus is on the trajectories that minimizes the
action and the generalized Lie symmetries of them. These trajectories
are solutions of the Euler-Lagrange equations of motion, and for almost
regular Lagrangians such solutions form a family of curves. It is, in fact, the
presence of this family of curves that gives rise to the generalized
Lie symmetry. The treatment here follows closely to that given in \cite{ADS2020}. 

For almost regular Lagrangians the solutions of the Euler-Lagrange
equations of motion 
\begin{equation}
M_{ab}(q, \dot{q})\ddot{q}^{b}=-\frac{\partial E}{\partial
  q^{a}}-F_{ab}(q, \dot{q})\dot{q}^{b}, 
\label{2ndEL1}%
\end{equation}
are not unique. While for these Lagrangians the rank of
$M_{ab}\left(q, \dot{q} \right) =D-N_{0}$\textemdash with $N_0=\hbox{dim
}\left(\hbox{ker }M_{ab}(q, \dot{q})\right)$\textemdash is constant, this rank
is not maximal, and thus Eq.~$(\ref{2ndEL1})$ does not have a unique
solution for $\ddot{q}$. Instead, for a chosen set of initial data
$\left(q_{0}=q(t_0),\dot{q}_{0}=\dot{q}(t_0)\right)$, the solution to
Eq.~$(\ref{2ndEL1})$ results in a family of solutions that evolve from
this $(q_0, \dot{q}_0)$. As with the paths in \textbf{Section \ref{&A-Sym}}, these
solutions are related to one another through a generalized Lie 
symmetry \cite{Olv1993}.

Following \cite{Olv1993}, the collection of functions
\begin{equation}
  \Delta_a(q,\dot{q}, \ddot{q}) := \frac{\partial E(q, \dot{q})}{\partial q^a} +
  F_{ab}(q, \dot{q})\dot{q}^b + M_{ab}(q, \dot{q}) \ddot{q}^b,  
\label{delta}
\end{equation}
defines a set of  surfaces $\Delta_a(q,\dot{q}, \ddot{q})=0$ on
$\mathbb{M}^{(2)}$, while the family of solutions to Eq.~$(\ref{2ndEL1})$ 
\begin{equation}
\mathcal{O}\left(q_0, \dot{q}_0\right)  :=\big\{q\left(  t\right)
\ \vert \ \
\Delta_a(q,\dot{q}, \ddot{q}) =0 \hbox{ with }
q\left(  t_{0}\right)
=q_{0},\ \dot{q}\left(  t_{0}\right)  =\dot{q}_{0}\big\}  ,
\nonumber
\end{equation}
that evolve from the same initial data $(q_0, \dot{q}_0)$ gives the collection
of trajectories that lie on these surfaces. Indeed, for any two such 
solutions $q^a(t)$ and $Q^a(t)$ there exists a 
$\mathfrak{z}(q, \dot{q})\in\hbox{ker } 
M_{ab}(q, \dot{q})$ such that $\ddot{Q}^a-\ddot{q}^a =
\mathfrak{z}^a$. Importantly, because $\mathfrak{z}^a$ depends on both $q$
and $\dot{q}$, the symmetry group that maps one member of
$\mathcal{O}$ to another must be a generalized Lie symmetry. We
therefore take the generator of this symmetry group to be 
\begin{equation}
  \mathbf{g} := \rho(q, \dot{q})\cdot \frac{\mathbf{\partial}
    \>\>\>}{\mathbf{\partial} q},
  \nonumber
\end{equation}
with the corresponding the second-order prolongation vector
for $\mathbf{g}$ being,
\begin{equation}
  \hbox{\textbf{pr }}\mathbf{g} := \rho \cdot
  \frac{\mathbf{\partial}\>\>\>}{\mathbf{\partial} q} 
  + \dot{\rho} \cdot \frac{\mathbf{\partial}\>\>\>}{\mathbf{\partial} \dot{q}} +
  \ddot{\rho}\cdot \frac{\mathbf{\partial} \>\>\>}{\mathbf{\partial} \ddot{q}},
  \nonumber
\end{equation}
with this $\hbox{\textbf{pr }}\mathbf{g}\in\mathbf{T}\mathbb{M}^{(2)}$. As with
the above, the total time 
derivative is given by Eq.~$(\ref{Dt})$, but unlike the analysis in
\textbf{Section \ref{&A-Sym}}, the evolution of the path\textemdash and indeed,
for all the trajectories in $\mathcal{O}(q_0,\dot{q}_0)$\textemdash here is
given by the Euler-Lagrange equations of motion.

The action of this
prolongation on $\Delta_a$ on the $\Delta_a=0$ surface gives,   
\begin{equation}
  \hbox{\textbf{pr }}\mathbf{g}\left[\Delta_a(q,
    \dot{q},\ddot{q})\right] = -\frac{\partial 
    \ddot{q}^b}{\partial q^a}M_{bc}(q, \dot{q})\rho^c +
  \frac{d\>\>\>}{dt}\left[F_{ab}(q, \dot{q})\rho^b + M_{ab}(q,
    \dot{q})\dot{\rho}^b\right]. 
  \nonumber   
\end{equation}
Since $N_0>0$, $\ddot{q}$ is not unique on this surface, and yet 
$\mathbf{g}$ must generate the same symmetry group for all the
trajectories in $\mathcal{O}(q_0, \dot{q}_0)$. Necessarily, 
$\rho(q, \dot{q})\in\hbox{ker }M_{ab}(q, \dot{q})$. It then follows
that $\hbox{\textbf{pr g}}[\Delta_a(q,\dot{q}, \ddot{q})] =0$
if and only if (iff) there are constants $b_a$ such that $b_a = 
F_{ab}\rho^b + M_{ab}\dot{\rho}^b$. The solutions in $\mathcal{O}(q_0,
\dot{q}_0)$ all have the same initial data, however, and thus
necessarily $\rho(q_0, \dot{q}_0)=0=\dot{\rho}(q_0,\dot{q}_0)$. We
conclude that $b_a=0$. The following result, first proved in
\cite{ADS2020}, then follows.  

\begin{lemma} \label{GS} If $\mathbf{g}$ is
  a generalized infinitesimal symmetry of $\Delta_a$, then
  $\rho^a(q, \dot{q})\in\hbox{ker } M_{ab}(q, \dot{q})$, and
  $\dot{\rho}^a(q, \dot{q})$ is a solution of
    \begin{equation}
    0=F_{ab}(q, \dot{q})\rho^b(q, \dot{q}) +
    M_{ab}(q, \dot{q})\dot{\rho}^b(q, \dot{q}).
    \label{sol}
   \end{equation}
\end{lemma}

As before, we denote the set of all vector fields $\mathbf{g}$ that satisfy \textbf{Lemma
  $\mathbf{\ref{GS}}$} by  $\mathfrak{g}$, while 
$\hbox{\textbf{pr }}\mathfrak{g} := \{\hbox{\textbf{pr
}}\mathbf{g}\ \vert \ \ \mathbf{g}\in \mathfrak{g}\}$ is the set of
their prolongations. Once again $\hbox{\textbf{pr }}\mathfrak{g}$ is
involutive, and the conditions under which $\hbox{\textbf{pr
}}\mathfrak{g}$ 
generates a generalized Lie symmetry group are given in
\cite{Olv1993}. Note, however, that while $\rho=0$ and
$\dot{\rho}=\mathfrak{z}$ for any $\mathfrak{z}\in\hbox{ker
}M_{ab}(q, \dot{q})$ is a solution of 
  Eq.~$(\ref{sol})$, we require that $\dot{\rho} =
  \mathbf{d}\rho/\mathbf{d}t$; these solutions cannot be generators of the
  generalized Lie symmetry. Next, if $\dot{\rho}$ is a solution of
  Eq.~$(\ref{sol})$, then 
$\dot{\rho}^a+\mathfrak{z}$ is a solution of Eq.~$(\ref{sol})$ as
  well, and thus these solutions are not unique. This, along with the
  previous observation, leads us to generators that are constructed from
  equivalence classes of prolongations. Finally, Eq.~$(\ref{delta})$
  gives for any $\mathfrak{z} \in\hbox{ker }M_{ab}(q, \dot{q})$,
\begin{equation}
  0=\mathfrak{z}^a\left(\frac{\partial E}{\partial
    q^a}+F_{ab}(q,\dot{q})\dot{q}^b\right),
  \label{e}
\end{equation}
on the solution surface $\Delta_a(q,\dot{q},\ddot{q})=0$. If
Eq.~(\ref{e}) does not hold identically, it must be
imposed, leading to Lagrangian constraints
\cite{Car1990a}. More importantly, because each
$q(t)\in\mathcal{O}(\mathfrak{u}_0)$   
must lie on the Lagrangian constraint submanifold, any symmetry transformation of
$q(t)$ generated by $\mathbf{pr }\>\> \mathbf{g}$ must give a path
$Q(t)$ that also lies on the constraint submanifold.

Not all vectors in $\mathbf{pr}\> \mathfrak{g}$ will be generators of
the generalized Lie symmetry group for
$\mathcal{O}(\mathfrak{u}_0)$. Determining which of these vectors are,
and the relationship between the generators of symmetries of
the Euler-Lagrange equations of motion and those of 
the action, is best done within the Lagrangian phase space
framework. To accomplish this, we will need the following generalization of
\textbf{Lemma \ref{GS}}.

Consider the vector
\begin{equation}
  \mathbf{k}:=c\cdot\frac{\mathbf{\partial}\>\>\>}{\mathbf{\partial} q}
  +\dot{c}\cdot\frac{\mathbf{\partial}\>\>\>}{\mathbf{\partial}
    \dot{q}},
    \nonumber
\end{equation}
with a $c\in\hbox{ker }M_{ab}(q, \dot{q})$ along with the quantity
\begin{equation}
  l_a:=F_{ab}c^b(q, \dot{q})+M_{ab}\dot{c}^b(q,\dot{q}).
  \nonumber
\end{equation}
After an integration by parts,
\begin{eqnarray}
    l_a&=& c^b(q,
    \dot{q})\left\{F_{ab}(q, \dot{q})-\frac{\mathbf{d}\>\>\,}{\mathbf{d}t}
    \frac{\mathbf{\partial}^2L}{\mathbf{\partial}
        \dot{q}^a\mathbf{\partial} \dot{q}^b}\right\},
      \nonumber
      \\
      &=& c^b(q,
    \dot{q})\left\{F_{ab}(q, \dot{q})-\left[\frac{\mathbf{d}\>\>\,}{\mathbf{d}t}, \frac{\partial
        \>\>\,}{\mathbf{\partial} \dot{q}^a}\right]\frac{\mathbf{\partial} L}{\mathbf{\partial}
      \dot{q}^b}- \frac{\mathbf{\partial}\>\>\,}{\mathbf{\partial}
      \dot{q}^a} \left(\frac{\mathbf{d}\>\>\,}{\mathbf{d}t}
    \frac{\mathbf{\partial} L}{\mathbf{\partial}
      \dot{q}^b}\right)\right\}. 
    \nonumber
\end{eqnarray}
Using Eq.~$(\ref{Dt})$ we have
\begin{equation}
  \left[\frac{\mathbf{d}\>\>\,}{\mathbf{d}t}, \frac{\mathbf{\partial}
      \>\>\,}{\mathbf{\partial} \dot{q}^a}\right]\frac{\partial L}{\partial \dot{q}^b} =
  -\frac{\mathbf{\partial}^2L}{\mathbf{\partial} q^a\mathbf{\partial}
    \dot{q}^b}-
  \frac{\mathbf{\partial}
    \ddot{q}^c}{\mathbf{\partial}
    \dot{q}^a}\frac{\mathbf{\partial}^2L}{\mathbf{\partial}
    \dot{q}^c\mathbf{\partial} \dot{q}^b}.
  \nonumber
\end{equation}
As $q(t)$ is  a solution of the Euler-Lagrange equations of
motion, we find that
\begin{equation}
l_a= c^b(q, \dot{q})\left\{F_{ab}(q, \dot{q})
+\frac{\mathbf{\partial}^2L}{\mathbf{\partial} q^a\mathbf{\partial}
    \dot{q}^b}-\frac{\mathbf{\partial}^2L}{\mathbf{\partial}
      \dot{q}^a\mathbf{\partial}q_b} +
  \frac{\mathbf{\partial}
    \ddot{q}^c}{\mathbf{\partial}
    \dot{q}^a}\frac{\mathbf{\partial}^2L}{\mathbf{\partial}
    \dot{q}^c\mathbf{\partial} \dot{q}^b}
  \right\}.
    \nonumber
\end{equation}
This last expression vanishes after the definition of $F_{ab}(q,
\dot{q})$ is used along with the requirement that $c\in\hbox{ker
}M_{ab}(q,\dot{q})$. We then have the following result. 

\begin{lemma} \label{AllS} For any vector 
\begin{equation}
  \mathbf{k}=c\cdot\frac{{\partial}\>\>\>}{{\partial} q}
  +\dot{c}\cdot\frac{{\partial}\>\>\>}{{\partial}
    \dot{q}},
  \nonumber
\end{equation}
such that $c\in\hbox{ker }M_{ab}$, 
\begin{equation}
  0=F_{ab}c^b(q, \dot{q})+M_{ab}\dot{c}^b(q,\dot{q}).
  \nonumber
\end{equation}.
\end{lemma}

\section{Generators of the Generalized Lie Symmetry for the
  Euler-Lagrange Equations of Motion\label{&review}}

The generators of the generalized Lie symmetry for both the
Euler-Lagrange equations of motion and the action are best found using
the Lagrangian phase space approach to mechanics. This phase space
and its concomitant mathematical structure provide the tools needed to
determine both the generators of the symmetry and the solutions to the
energy equation on which they act. For the Euler-Lagrange equations of 
motion this determination was done in \cite{ADS2020}. In this
section we will review the Lagrangian phase space approach, establish
the notation used in this paper, and summarize the results obtained in
\cite{ADS2020} that are needed here. (We will also take the
opportunity to correct typographical errors made in \cite{ADS2020}.)
Proofs of the majority of the assertions listed in this section will
not be given; the reader is instead referred to \cite{ADS2020} where the
proofs and the context of their development can be found.

\subsection{The Lagrangian Phase space\label{&phase}}

For a configuration space $\mathbb{Q}$ the \textbf{Lagrangian phase
  space} $\mathbb{P}_L$ is the tangent space
$\mathbb{P}_L=\mathbf{T}\mathbb{Q}$, with the coordinates on
$\mathbb{P}_L$ denoted as $\mathfrak{u}=(q^1, \dots, q^D, v^1, \dots 
v^D)$. Integral flows on $\mathbb{P}_L$,
$t\in[t_0,\infty)\to\mathfrak{u}(t)\in\mathbb{P}_L$ 
\cite{Abr1978}, for a set of initial data $\mathfrak{u}_0=(q_0,
v_0)$ are given as solutions to  
\begin{equation}
\frac{d\mathfrak{u}}{dt}:=\mathbf{X}  (\mathfrak{u}),
\nonumber
\end{equation}
where $\mathbf{X}$ is a smooth vector field in 
$\mathbf{T}\mathbb{P}_L=\mathbf{T}(\mathbf{T}\mathbb{Q})$. The two
tangent spaces $\mathbf{T}\mathbb{Q}$ and $\mathbf{T}\mathbb{P}_L$
have the bundle projections: $\tau_{\mathbb{Q}}:\mathbf{T}\mathbb{Q}\to  
\mathbb{Q}$ and 
$\tau_{\mathbf{T}\mathbb{Q}}:\mathbf{T}(\mathbf{T}\mathbb{Q})
\to\mathbf{T}\mathbb{Q}$. They can be used to
construct two other projection maps: $\tau_{\mathbb{Q}}\circ 
\tau_{\mathbf{T}\mathbb{Q}}:\mathbf{T}(\mathbf{T}\mathbb{Q})  
\to\mathbb{Q}$ and the prolongation of $\tau_{\mathbf{T}\mathbb{Q}}$
to $\mathbf{T}(\mathbf{T}\mathbb{Q})$ (see \cite{Got1979} and
\cite{Abr1978}). This prolongation is the map 
$\mathbf{T}\tau_{\mathbb{Q}}:\mathbf{T}(\mathbf{T}\mathbb{Q})\to\mathbf{T}\mathbb{Q}$,
and is defined by requiring that the two maps
$\tau_{\mathbb{Q}}\circ \tau_{\mathbf{T}\mathbb{Q}}$ and 
$\tau_{\mathbb{Q}}\circ \mathbf{T}\tau_{\mathbb{Q}}$ map any point in
$\mathbf{T}(\mathbf{T}\mathbb{Q})$ to the same point in
  $\mathbb{Q}$. The \textbf{vertical subbundle}
$[\mathbf{T}\mathbb{P}_L]^v$ of $\mathbf{T}(\mathbf{T}\mathbb{Q})$
is $[\mathbf{T}\mathbb{P}_L]^v = \hbox{ker
}\mathbf{T}\tau_{\mathbb{Q}}$ \cite{Got1979}; a $\mathbf{X}^v\in
[\mathbf{T}_{\mathfrak{u}}\mathbb{P}_L]^v$ above a point
$\mathfrak{u}\in\mathbb{P}_L$ is called a \textbf{vertical vector
  field}. The \textbf{horizontal subbundle} 
$[\mathbf{T}\mathbb{P}_L]^q$ of
$\mathbf{T}(\mathbf{T}\mathbb{Q})$ is 
$[\mathbf{T}\mathbb{P}_L]^q = \hbox{Image
}\mathbf{T}\tau_{\mathbb{Q}}$; a
$\mathbf{X}^q\in[\mathbf{T}_{\mathfrak{u}}\mathbb{P}_L]^q$ is called a 
\textbf{horizontal vector field}. Consequently, each  
$\mathbf{X}\in\mathbf{T}_{\mathfrak{u}}\mathbb{P}_L$ consists of a
$\mathbf{X}^q \in
\left[\mathbf{T}_{\mathfrak{u}}\mathbb{P}_L\right]^q$ and a
$\mathbf{X}^v \in
\left[\mathbf{T}_{\mathfrak{u}}\mathbb{P}_L\right]^v$ with $\mathbf{X}
= \mathbf{X}^q + \mathbf{X}^v$. In terms of local coordinates,  
\begin{equation}
\mathbf{X}^{q}
:=X^{qa}\frac{\mathbf{\partial}\>\>\>}{\mathbf{\partial}q^{a}} ,\quad
\hbox{and}\quad\mathbf{X}^{v} :=X^{va}
\frac{\mathbf{\partial}\>\>\>}{\mathbf{\partial}v^{a}}.
\nonumber
\end{equation}
Of special interest is the second order Lagrangian vector field
$\mathbf{X}_L$. This vector field is the particular solution of
Eq.~$(\ref{EnergyE})$ for which 
$\mathbf{T}\tau_{\mathbb{Q}}\circ\mathbf{X}_L$ is the identity on
$\mathbf{T}\mathbb{Q}$ (see \cite{Abr1978}). In terms of local
coordinates
\begin{equation}
  \mathbf{X}_L = v^a \frac{\mathbf{\partial}\>\>\>}{\mathbf{\partial} q^a} +
  X^{va} \frac{\mathbf{\partial}\>\>\>}{\mathbf{\partial}v^a}.
  \nonumber
\end{equation}

The space of one-forms on
$\mathbb{P}_L$ is the cotangent space
$\mathbf{T}^{*}\mathbb{P}_L$. For a one-form $\mathbf{\alpha}\in 
\mathbf{T}^{*}_{\mathfrak{u}}\mathbb{P}_L$, and a vector
field $\mathbf{X}\in\mathbf{T}_{\mathfrak{u}}\mathbb{P}_L$, the
dual prolongation map $\mathbf{T}^{*}\tau_{\mathbb{Q}}$ is defined as
\begin{equation}
  \langle \mathbf{\alpha}\vert
  \mathbf{T}\tau_{\mathbb{Q}}\mathbf{X}\rangle = \langle
  \mathbf{T}^{*}\tau_{\mathbb{Q}} \mathbf{\alpha}\vert
  \mathbf{X}\rangle,
  \nonumber
\end{equation}
after a useful adaptation of Dirac's bra and ket notation. In
addition, for a general $k$-form \textbf{$\mathbf{\omega}$} in the
$k$-form bundle $\mathbf{\Lambda}^{k}\left(\mathbb{P}_L\right)$, 
\begin{equation}
\mathbf{\omega}\left(  x\right)  :\mathbf{Y}_{1}\otimes\cdots\otimes
\mathbf{Y}_{k}\rightarrow\left\langle \left.  \mathbf{\omega}\left(  x\right)
\right\vert \mathbf{Y}_{1}\otimes\cdots\otimes\mathbf{Y}_{k}\right\rangle
\in\mathbb{R},
\nonumber
\end{equation}
with
$\mathbf{Y}_{j}\in\mathbf{T}_{\mathfrak{u}}\mathbb{P}_L$ for $j = 1,
\dots, k$. The \textbf{vertical one-form subbundle}
$[\mathbf{T}^{*}\mathbb{P}_L]^v$ of $\mathbf{T}^{*}\mathbb{P}_L$ is 
$[\mathbf{T}^{*}\mathbb{P}_L]^v :=\hbox{ker }
\mathbf{T}^{*}\tau_{\mathbb{Q}}$; a
$\mathbf{\alpha}_v\in [\mathbf{T}^{*}_{\mathfrak{u}}\mathbb{P}_L]^v$ is
called a \textbf{vertical one-form}. The \textbf{horizontal
one-form subbundle} $[\mathbf{T}^{*}\mathbb{P}_L]^q$ of
$\mathbf{T}^{*}\mathbb{P}_L$ is 
$[\mathbf{T}^{*}\mathbb{P}_L]^q=\hbox{Image }
\mathbf{T}^{*}\mathbb{P}_L$; a
$\mathbf{\alpha}_q\in[\mathbf{T}^{*}_{\mathfrak{u}}\mathbb{Q}]^q$ is
called a \textbf{horizontal one-form}. Each one-form $\mathbf{\varphi}\in
\mathbf{T}^{*}_{\mathfrak{u}}\mathbb{P}_L$ consists of a
$\mathbf{\varphi}_{q} \in \left[\mathbf{T}_{\mathfrak{u}}^{*}\mathbb{P}_L\right]^q$ and
a $\mathbf{\varphi}_{v} \in 
\left[\mathbf{T}_{\mathfrak{u}}^{*}\mathbb{P}_L\right]^q$
such that $\mathbf{\varphi}=\mathbf{\varphi}_{q}
+\mathbf{\varphi}_{v}$. In terms
of local coordinates 
$\mathbf{\varphi}_{q} :=\varphi_{qa} \ \mathbf{d}q^{a}$ and 
$\mathbf{\varphi}_{v} :=\varphi_{va} \mathbf{d}v^{a}$. 

Following \cite{Got1979, Got1980}, the Lagrangian two-form is defined
as $\mathbf{\Omega}_L := -\mathbf{d}\mathbf{d}_JL$, 
where $\mathbf{d}_J$ is the vertical derivative (see
\cite{Got1979}). This two-form can be expressed as
$\mathbf{\Omega}_{L}:=\mathbf{\Omega}_{F}+\mathbf{\Omega}_{M}$ such
that for any $\mathbf{X},
\mathbf{Y}\in\mathbf{T}_{\mathfrak{u}}\mathbb{P}_L$. 
\begin{equation}
\mathbf{\Omega}_F(\mathbf{X},\mathbf{Y}) :=
\mathbf{\Omega}_L(\mathbf{T}\tau_{\mathbb{Q}}\mathbf{X},
\mathbf{T}\tau_{\mathbb{Q}}\mathbf{Y}),
\nonumber
\end{equation}
and is thus the \textbf{horizontal two-form} of $\mathbf{\Omega}_L$. As
$\mathbf{\Omega}_M(\mathbf{X},\mathbf{Y})=\mathbf{\Omega}_L(\mathbf{X},\mathbf{Y})- 
\mathbf{\Omega}_F(\mathbf{X},\mathbf{Y})$, $\mathbf{\Omega}_M$ is then
a \textbf{mixed two-form} of 
$\mathbf{\Omega}_L$. In terms of local coordinates,  
\begin{equation}
\mathbf{\Omega}_{L}=-\mathbf{d{\theta}}_{L},\quad \hbox{where} \quad
\mathbf{\theta}_{L}:=\frac{\partial L}{\partial v^{a}%
}\mathbf{d}q^{a},
\nonumber
\end{equation}
while
\begin{equation}
\mathbf{\Omega
}_{F}:=\frac{1}{2}F_{ab}\mathbf{d}q^{a}\wedge\mathbf{d}q^{b}%
,\ \hbox{and} \ \mathbf{\Omega}_{M}:=M_{ab}\mathbf{d}q^{a}\wedge\mathbf{d}v^{b}.
\nonumber
\end{equation}

For regular Lagrangians $\mathbf{X}_L$ is the unique solution of
Eq.~$(\ref{EnergyE})$. For almost regular Lagrangians, on the other
hand, this solution is not unique, but instead depends on 
\begin{equation}
\ker\>  \mathbf{\Omega}_{L}\left(  \mathfrak{u}\right) 
:=\left\{  \mathbf{K}\in\mathbf{T}_{\mathfrak{u}}\mathbb{P}_{L}%
\ \vert\ \ i_{\mathbf{K}}\mathbf{\Omega}_{L}=0\right\}.
\nonumber
\end{equation}
From \textbf{Section \ref{&Sym}} we expect this kernel to
play a role in determining the generators of the generalized Lie 
symmetry of both the Euler-Lagrange equations of motion and the
action. Indeed, consider the natural isomorphism $iso: (t, q, \dot{q}, 
\ddot{q}) \in \mathbb{M}^{(2)} \to (t, q, v, X^{va}_{L})$ defined in
\cite{ADS2020}, and the prolongation $\hbox{\textbf{pr} }\mathbf{g}$ 
of a generator $\mathbf{g}\in\mathbf{\mathfrak{g}}$ of a 
generalized Lie symmetry of the Euler-Lagrange equations of
motion. This $\hbox{\textbf{pr }}\mathbf{g}$ contains the vector  
\begin{equation}
  \mathbf{k}=\rho\cdot\frac{\mathbf{\partial} \>\>\>}{\mathbf{\partial} q} +
  \dot{\rho}\cdot\frac{\mathbf{\partial} \>\>\>}{\mathbf{\partial}\dot{q}}.
  \nonumber
\end{equation}
The collection of all such vectors has been shown to be
involutive (see \cite{ADS2020}). The isomorphism maps $iso:
\mathbf{k}\to\mathbf{k}'$ where 
\begin{equation}
  \mathbf{k}'=\rho\cdot\frac{\mathbf{\partial} \>\>\>}{\mathbf{\partial} q} +
  \dot{\rho}\cdot\frac{\mathbf{\partial} \>\>\>}{\mathbf{\partial} v}.
  \nonumber
\end{equation}
Then $\mathbf{k}'\in\mathbf{T}\mathbb{P}_L$, and from
\textbf{Lemma \ref{GS}}, $\mathbf{k}'\in\hbox{ker
}\mathbf{\Omega}_L(\mathfrak{u})$ as well. A similar result holds for
the generators in $\mathbf{\mathfrak{g}}_L$ after 
\textbf{Lemma \ref{Action-Sym}} and \textbf{Lemma \ref{AllS}} are used.

The two-form $\mathbf{\Omega}_L$ gives the lowering map
$\mathbf{\Omega}_L^{\flat}:\mathbf{T}_{\mathfrak{u}}\mathbb{P}_{L}\rightarrow 
\mathbf{T}_{\mathfrak{u}}^{\ast}\mathbb{P}_{L}$, with
$\Omega_L^{\flat}\mathbf{X}:=i_{\mathbf{X}}\mathbf{\Omega}_L$.
This map consists of
$\Omega_{L}^{\flat}=\Omega_{F}^{\flat}+\Omega_{M}^{v\flat}+\Omega_{M}^{q\flat}$, 
with $\Omega_{F}^{\flat}:
\mathbf{X}\in\mathbf{T}_{\mathfrak{u}}\mathbb{P}_L \to
\left[\mathbf{T}^{*}_{\mathfrak{u}}\mathbb{P}_L\right]^q$; 
$\Omega_{M}^{q\flat}:\mathbf{X}\in\mathbf{T}_{\mathfrak{u}}\mathbb{P}_L 
\to\left[\mathbf{T}^{*}_{\mathfrak{u}}\mathbb{P}_L\right]^q$; and
$\Omega_{M}^{v\flat}:
\mathbf{X}\in\mathbf{T}_{\mathfrak{u}}\mathbb{P}_L \to
\left[\mathbf{T}^{*}_{\mathfrak{u}}\mathbb{P}_L\right]^v$. In terms of local
coordinates, 
$\Omega_{F}^{\flat}\mathbf{X} = F_{ab}X^{qa}\mathbf{d}q^b$, 
$\Omega_{M}^{q\flat}\mathbf{X}= -M_{ab}X^{va}{}\mathbf{d}q^{b}$, and
$\Omega_{M}^{v\flat}\mathbf{X}= M_{ab}X^{qa}\mathbf{d}v^{b}$.

For almost regular Lagrangians $\ker\>
  \Omega_{M}^{v\flat} = \mathcal{C}\oplus
  \left[ \mathbf{T}_{\mathfrak{u}} \mathbb{P}_{L}\right]  ^{v}$ 
  while $\ker\>  \Omega_{M}^{q\flat}  =\left[
    \mathbf{T}_{\mathfrak{u}}\mathbb{P}_{L}\right]  ^{q}\oplus  
  \mathcal{G}$. Here
  \begin{equation}
    \mathcal{C}:=\left\{\mathbf{C}\in[\mathbf{T}_q\mathbb{P}_L]^q
      \ \vert \ i_{\mathbf{C}}\mathbf{\Omega}_M =0\right\},
      \nonumber
  \end{equation}
  and
  \begin{equation}
    \mathcal{G}:=\left\{\mathbf{G}\in
    [\mathbf{T}_q\mathbb{P}_L]^v \ \vert
    \ i_{\mathbf{G}}\mathbf{\Omega}_M =0\right\}.
    \nonumber
  \end{equation}
As $M_{ab}(\mathfrak{u})$ has constant rank on
$\mathbb{P}_L$, there exists a basis,  
\begin{equation}
\Big\{
\mathbf{\mathfrak{z}}_{\left(  n\right)  }\left(  \mathfrak{u}\right) =\left(  \mathfrak{z}%
_{\left(  n\right)  }^{1}\left(  \mathfrak{u}\right)  ,\ldots,\mathfrak{z}%
_{\left(  n\right)  }^{D}\left(  \mathfrak{u}\right)  \right)  \vert
\ M_{ab}\left(  \mathfrak{u}\right)  \mathfrak{z}_{\left(  n\right)  }%
^{b}\left(  \mathfrak{u}\right)  =0,
n=1,\ldots,N_{0} \Big\}  ,
\nonumber
\end{equation}
for $\ker  M_{ab}\left(  \mathfrak{u}\right) $ at each
$\mathfrak{u}\in\mathbb{P}_{L}$. Spans of both
\begin{eqnarray}
  \mathcal{C}&=& \hbox{span }
  \left\{\mathbf{U}^q_{(n)}=\mathbf{\mathfrak{z}}_{(n)}\cdot
    \frac{\mathbf{\partial}\>\>\>}{\mathbf{\partial} q}, n=1,
    \dots, N_0\right\}, \hbox{and}\>
    \nonumber
    \\
  \mathcal{G}&=& \hbox{span }
  \left\{\mathbf{U}^v_{(n)}=\mathbf{\mathfrak{z}}_{(n)}\cdot
    \frac{\mathbf{\partial}\>\>\>}{\mathbf{\partial} v}, n=1,
    \dots, N_0\right\},
    \nonumber
\end{eqnarray}
can then be constructed. Importantly, $\mathcal{G}$ is involutive
\cite{Car1990a}, and when the rank of
$\mathbf{\Omega}_L(\mathfrak{u})$ is constant on $\mathbb{P}_L$,
$\hbox{ker } \mathbf{\Omega}_L(\mathfrak{u})$ is involutive as well.

Corresponding to $\mathbf{U}^q_{(n)}$ and $\mathbf{U}^v_{(n)}$ we have
the one-forms  
$\mathbf{\Theta}^{(m)}_q$ and $\mathbf{\Theta}^{(m)}_v$
where $\langle\mathbf{\Theta}^{(m)}_q\vert
\mathbf{U}^q_{(n)} \rangle= \delta_{(n)}^{(m)}$ and
$\langle\mathbf{\Theta}^{(m)}_v\vert \mathbf{U}^v_{(n)} \rangle=
\delta_{(n)}^{(m)}$. Then $\left[  \mathbf{T}%
_{\mathfrak{u}}\mathbb{P}_{L}\right]  ^{q}=%
\mathcal{C}%
\oplus%
\mathcal{C}%
_{\perp}$ and $\left[  \mathbf{T}_{\mathfrak{u}}\mathbb{P}_{L}\right]  ^{v}=%
\mathcal{G}%
\oplus%
\mathcal{G}%
_{\perp}$, where 
\begin{eqnarray}%
\mathcal{C}%
_{\perp}:=\bigg\{  \mathbf{X}\in\left[  \mathbf{T}_{\mathfrak{u}}%
  \mathbb{P}_{L}\right]  ^{q}\ &\vert&\
\left\langle \left.  \mathbf{\Theta}_{q}^{(n)}\right\vert \mathbf{X}%
\right\rangle =0,\>\>  n=1,\dots,N_{0}  \ \bigg\}  ,
\hbox{and}%
\nonumber
\\
\mathcal{G}_{\perp}:=\bigg\{  \mathbf{X}\in\left[
  \mathbf{T}_{\mathfrak{u}}
  \mathbb{P}_{L}\right]  ^{v}\ &\vert&\
\left\langle \left.  \mathbf{\Theta}_{v}^{(n)}\right\vert \mathbf{X}%
\right\rangle =0, \>\>   n=1,\dots,N_{0}  \ \bigg\}.
\nonumber
\end{eqnarray}
The vectors that lie in $\hbox{ker
}\mathbf{\Omega}_L(\mathfrak{u})$ can be determined by using the reduced
matrix $ {F}_{nm}:=\mathfrak{z}_{\left(  n\right)
}^{a}F_{ab}\mathfrak{z}_{\left( 
m\right)  }^{b}$ to define
\begin{equation}
\overline{%
\mathcal{C}%
}:=\left\{  \overline{\mathbf{C}}\in%
\mathcal{C}%
\ \bigg\vert\ \sum_{m=1}^{N_{0}}\bar{F}_{nm}\overline{C}^{\left(  m\right)
}=0\right\}  \subset%
\mathcal{C}.%
\nonumber
\end{equation}
Then,

\begin{theorem}
\label{@NVGen}The vectors $\mathbf{K=K}^{q}+\mathbf{K}^{v}\in\ker
\mathbf{\Omega}_{L}$ are given by,%
\begin{equation}
\mathbf{K}^{q}=\overline{\mathbf{C}}\mathbf{,\ \ K}^{v}=\mathbf{G+}%
\widehat{\mathbf{C}},
\nonumber
\end{equation}
where, $\overline{\mathbf{C}}\in\overline{%
\mathcal{C}%
}$, $\mathbf{G}\in%
\mathcal{G}%
$, and $\widehat{\mathbf{C}}\in%
\mathcal{G}%
_{\perp}$ is the unique solution of $M_{ab}\widehat{C}^{b}=-F_{ab}\overline
{C}^{b}$.
\end{theorem}
We found in \cite{ADS2020} that $\dim\>  \left(\ker
\mathbf{\Omega}_{L}\left(  \mathfrak{u}\right)\right) 
  =N_{0}+\bar{D}, $ where $
\ \bar{D}  :=\dim \>\overline{\mathcal{C}}\le N_0$ (see \cite{ADS2020}
for proof). However, the results of \textbf{Lemma \ref{AllS}} show
that we can construct from any vector $\mathbf{U}^q\in \mathcal{C}$ a
vector that lies in the $\hbox{ker }\mathbf{\Omega}_L(\mathfrak{u})$,
and as dim $(\mathcal{C})=N_0$, it follows that dim $(\hbox{ker
}\mathbf{\Omega}_L(\mathfrak{u}))=2N_0$. 

\subsection{First-order Lagrangian constraints\label{&LCon}}

For singular Lagrangians solutions of the energy equation
$\mathbf{X}_E$ are not unique. It is well known that they also do not,
in general, exist throughout $\mathbb{P}_L$, but are instead confined
to a submanifold of the space given by Lagrangian constraints. 

With $\mathbf{X}_{E}=\mathbf{X}^q_{E}+\mathbf{X}^v_{E}$, it is
convenient to use the one form $\mathbf{\Psi}$
\begin{equation}
  \Omega_{M}^{q\flat}\mathbf{X}_{E}^{v}=\mathbf{\Psi}.
\nonumber
\end{equation}
constructed from the energy equation. The \textbf{first-order
  constraint functions} are then $\gamma_{n}^{\left[
    1\right]  }:=\left\langle 
\left.  \mathbf{\Psi}\right\vert \mathbf{U}_{\left(  n\right)
}^{q}\right\rangle=0$ for $n=1,\ldots,N_{0}$. In terms of local coordinates,
\begin{equation}
\gamma_{n}^{\left[  1\right]  }= U_{\left(  n\right)  }%
^{qa}\left(  \frac{\partial E}{\partial q^{a}}+F_{ab}v^{b}\right).
\nonumber
\end{equation}
They may also be expressed \cite{Got1979,
  Got1980} as $\gamma^{[1]}_n = \langle
\mathbf{d}E\vert\mathbf{P}_{(n)}\rangle=\mathbf{P}_{(n)}E$ for any
basis $\{\mathbf{P}_{(n)}\}$ of ker $\mathbf{\Omega}_L(\mathfrak{u})$
for which $\langle\mathbf{\Theta}^{(m)}_q\vert
\mathbf{P}_{(n)}\rangle=\delta^{(m)}_{(n)}$. 
In general, $\gamma_{n}^{\left[  1\right]  }\ne0$ on
$\mathbb{P}_L$. Instead, the condition $\gamma_{n}^{\left[  1\right]
}=0$ must be imposed, and this in turn defines a set of submanifolds
of $\mathbb{P}_L$ given by the collection $\hbox{C}_{L}^{\left[  1\right]  }:=\left\{
\gamma_{1}^{\left[  1\right]  },\ldots,\gamma_{N_{0}}^{\left[  1\right]
}\right\}  $. The collection of these
surfaces, 
$\mathbb{P}_{L}^{\left[  1\right]  }:=\left\{ 
\mathfrak{u}\in\mathbb{P}_{L}\ \vert\ \ \gamma_{n}^{\left[  1\right]
}\left(  \mathfrak{u}\right)  =0\, ,  n=1,\ldots,N_{0}\ \right\}$ is
called the \textbf{first-order Lagrangian constraint submanifold}, and
has $\dim \mathbb{P}_{L}^{\left[1\right]  } =2D-I_{\left[
    1\right]}$. Here $I_{\left[1\right]  }$ is the
number of independent functions in $\hbox{C}_{L}^{\left[  1\right]
}$ with $I_{\left[ 1\right]  }=\hbox{rank }  \left\{\mathbf{d}\gamma_{n}^{\left[
      1\right]  }\right\} \leq N_{0}$. 

The \textbf{constraint
  one-form}  
\begin{equation}
  \mathbf{\beta}[\mathbf{X}_E] :=
  \mathbf{d}E-i_{\mathbf{X}_E}\mathbf{\Omega}_L,
\nonumber
\end{equation}
was introduced in \cite{ADS2020} with the condition 
$\mathbf{\beta}[\mathbf{X}_E]=0$ giving both the  
solution of the energy equation and the submanifold
$\mathbb{P}_L^{[1]}$. As $\langle\mathbf{\beta} \vert \mathbf{U}^q_{(n)}\rangle=
\gamma_n^{[1]}$, this $\mathbf{\beta}[\mathbf{X}_E]$ can also
be expressed as 
\begin{equation}
  \mathbf{\beta}[\mathbf{X}_E]=\sum_{n=1}^{N_0}
  \gamma_n^{[1]}\mathbf{\Theta}^{(n)}_q.
  \label{beta}
\end{equation}

\subsection{The Generalized Lie Symmetry Group for the Euler-Lagrange
  Equations of motion\label{&GenEL-Sym}} 

The generalized Lie symmetry group for $\mathcal{O}(\mathfrak{u}_0)$
is determined using
\begin{equation}
  \overline{\hbox{ker }\mathbf{\Omega}_L(\mathfrak{u})} :=
  \{\mathbf{P} \in \hbox{ker } \mathbf{\Omega}_L(\mathfrak{u})\
  \vert \ [\mathbf{G},\mathbf{P}]
  \in\left[\mathbf{T}_{\mathfrak{u}}\mathbb{P}_L\right]^v \>\>\forall
  \>\>\mathbf{G}\in\mathcal{G}\},  
\end{equation}
along with the following collection of functions on $\mathbb{P}_L$,  
\begin{equation}
  \overline{\mathcal{F}} := \{f\in C^\infty \hbox{on } \mathbb{P}_L
  \ \vert\ \ \mathbf{G}f = 0 \>\>\forall\>\> \mathbf{G} \in
  \mathcal{G}\}.
\nonumber
\end{equation}
This $\overline{\hbox{ker }\mathbf{\Omega}_L(\mathfrak{u})}$ is also
involutive.

The following results were proved in \cite{ADS2020}.

\begin{lemma} \label{basic} Let  
  $\mathbf{X}\in\mathbf{T}_{\mathfrak{u}}\mathbb{P}_L$ 
  and $\mathbf{G}\in\mathcal{G}$ such that $[\mathbf{G},
    \mathbf{X}]\in\hbox{ker }\mathbf{\Omega}_L(\mathfrak{u})$. Then 
  $[\mathbf{G},
    \mathbf{X}]\in\left[\mathbf{T}_{\mathfrak{u}}\mathbb{P}_L\right]^v$
  iff $[\mathbf{G}, \mathbf{X}]\in\mathcal{G}$.
\end{lemma}
It then follows that $[\mathbf{G},\mathbf{P}]\in\mathcal{G}$ for all
$\mathbf{P}\in \overline{\hbox{ker
  }\mathbf{\Omega}_L(\mathfrak{u})}$. As $\mathcal{G}$ is involutive
and as $\mathcal{G}\subset \hbox{ker }\mathbf{\Omega}_L(\mathfrak{u})$,
$\mathcal{G}\subset \overline{\hbox{ker
  }\mathbf{\Omega}_L(\mathfrak{u})}$ as well, and thus   
$\mathcal{G}$ is an ideal of $\overline{\hbox{ker
  }\mathbf{\Omega}_L(\mathfrak{u})}$.  

\begin{lemma}
  \label{basis}
  There exists a choice of basis for ker
  $\mathbf{\Omega}_L(\mathfrak{u})$ that is 
  also a basis of $\overline{\hbox{ker
    }\mathbf{\Omega}_L(\mathfrak{u})}$.
\end{lemma}

As $\mathcal{G}$ is an ideal of $\overline{\hbox{ker
  }\mathbf{\Omega}_L(\mathfrak{u})}$, we may define for
any $\mathbf{P}_1, \mathbf{P}_2\in\overline{\hbox{ker 
  }\mathbf{\Omega}_L(\mathfrak{u})}$ the equivalence relation:
$\mathbf{P}_1\sim\mathbf{P}_2$ iff
$\mathbf{P}_1-\mathbf{P}_2\in\mathcal{G}$. The equivalence class,
\begin{equation}
  \left[\mathbf{P}\right]:=
  \{\mathbf{Y}\in \overline{\hbox{ker
    }\mathbf{\Omega}_L(\mathfrak{u})}\ \vert 
  \ \mathbf{Y}\sim\mathbf{P}\},  
\end{equation}
can be constructed along with the quotient space
$\overline{\hbox{ker }\mathbf{\Omega}_L(\mathfrak{u})}/\mathcal{G}$.   
(For the sake of notational clarity  we will suppress
the square brackets for equivalence classes when there
is no risk of confusion.) This space is a collection of
vectors that lie in the kernel of $\mathbf{\Omega}_L$, but with the
vectors in $\mathcal{G}$ removed; $\overline{\hbox{ker
  }\mathbf{\Omega}_L(\mathfrak{u})}/\mathcal{G}$ thereby addresses the
first two observations listed at the end of \textbf{Section
  \ref{&EL-Sym}}. 

We now turn our attention to the third observation. Because the
integral flow $\mathfrak{u}_{\mathbf{X}}(t)$ of any solution 
$\mathbf{X}$ of the energy equation must lie on $\mathbb{P}_L^{[1]}$, a 
symmetry transformation of $\mathfrak{u}_{\mathbf{X}}(t)$ must result
in an integral flow $\mathfrak{u}_{\mathbf{Y}}(t)$ of another solution
$\mathbf{Y}$ of the energy equation, which must also lie on
$\mathbb{P}_L^{[1]}$. Implementing this condition is done through
$\mathbf{\beta}[\mathbf{X}_E]$. 

As $\langle\mathbf{\beta}[\mathbf{X}_E]\vert \mathbf{G}\rangle=\langle
\mathbf{d}E\vert\mathbf{G}\rangle= \mathbf{G}E=0$ for all  
$\mathbf{G}\in\mathcal{G}$ on $\mathbb{P}_L^{[1]}$, the Lie derivative
$\mathfrak{L}_{\mathbf{G}}$ of $\mathbf{\beta}$ along $\mathbf{G}$ is, 
\begin{equation}
  \mathfrak{L}_{\mathbf{G}}\mathbf{\beta}[\mathbf{X}_E]= \sum_{n=1}^{N_0}
  \left(\mathbf{G}\gamma_n^{[1]}\right) \mathbf{\Theta}^{(n)}_q.
  \nonumber
\end{equation}
Given a
$\mathbf{P}_{(n)} \in\overline{\hbox{ker
  }\mathbf{\Omega}_L(\mathfrak{u})}$ such that $\mathbf{P}_{(n)} = 
\mathbf{U}_{(n)}^q + \widehat{\mathbf{U}}_{(n)} +\mathbf{G}'$ with
$\mathbf{G}'\in\mathcal{G}$, $\mathbf{G}\gamma_n^{[1]}=
       [\mathbf{G}, \mathbf{P}_{(n)}]E + 
\mathbf{P}_{(n)}\mathbf{G}E$. But $\mathcal{G}$ is an ideal of
$\overline{\hbox{ker }\mathbf{\Omega}_L(\mathfrak{u})}$, and thus
$\mathbf{G}\gamma_{n}^{[1]}=0$ on the first-order constraint
manifold. It follows that $\mathfrak{L}_{\mathbf{G}}\mathbf{\beta}
=0$ on $\mathbb{P}^{[1]}_L$. The collection of vectors,     
\begin{equation}
  \mathcal{S}\hbox{ym} := \big\{\mathbf{P}\in \overline{\hbox{ker
    }\mathbf{\Omega}_L(\mathfrak{u})}/\mathcal{G}
  \ \vert\ \
  \mathfrak{L}_{\mathbf{P}}\mathbf{\beta}[\mathbf{X}_E] =
  \mathbf{d}\langle \mathbf{\beta}[\mathbf{X}_E]\vert \mathbf{P}\rangle \hbox{ on }
  \mathbb{P}_L^{[1]}\big\}, 
  \nonumber
\end{equation}
is therefore well defined, and is involutive. It follows that
$\mathbf{P}\in\mathcal{S}\hbox{ym}$ iff    
$\langle \mathbf{d}\mathbf{\beta}[\mathbf{X}_E]\vert \mathbf{P}\otimes
\mathbf{X}\rangle=0$ for all $\mathbf{X}\in
\mathbf{T}\mathcal{P}_L$. We are then able to construct from each
$\mathbf{P}\in\mathcal{S}\hbox{ym}$ a one-parameter subgroup
$\mathbf{\sigma}_{\mathbf{P}}(\epsilon,x)$ defined as the solution
to  
\begin{equation}
  \frac{d\mathbf{\sigma}_{\mathbf{P}}}{d\epsilon} :=
  \mathbf{P}\left(\mathbf{\sigma}_{\mathbf{P}}\right), 
  \nonumber
\end{equation}
where $\sigma_{\mathbf{P}}(0,\mathfrak{u}) = \mathfrak{u}$ for
$\mathfrak{u}\in\mathbb{P}_L$. The collection of such 
subgroups with give the Lie group
$\hbox{Gr}_{\mathcal{S}\hbox{ym}}$.

\subsection{Euler-Lagrange Solutions of the Energy
  Equation\label{&Sol}}

We denote the set of \textbf{general solutions} to the energy equation as
\begin{equation}
\mathcal{S}\hbox{ol}
:=\{\mathbf{X}_{E}\in\mathbf{T}_{\mathfrak{u}}\mathbb{P}_L
\ \vert\ \ i_{\mathbf{X}_{E}}\mathbf{\Omega}_L = \mathbf{d}E
\hbox{ on } \mathbb{P}_L^{[1]}\}.
\nonumber
\end{equation}
If $\mathfrak{u}(t)$ is the integral flow of a vector in
$\mathcal{S}\hbox{ol}$ whose projection onto
$\mathbb{Q}$ corresponds to a solution of
the Euler-Lagrange equations of motion, then
$\hbox{Gr}_{\mathcal{S}\hbox{ym}}$ must map one of
such flows into another one. However, while
$\mathfrak{L}_{\mathbf{G}}\mathbf{X}_{L} = 
[\mathbf{G}, \mathbf{X}_{L}]\in \hbox{ker
}\mathbf{\Omega}_L(\mathfrak{u})$, in general
$\mathfrak{L}_{\mathbf{G}}\mathbf{X}_{L} \notin \mathcal{G}$. The 
action of $\sigma_{\mathbf{P}}$ on the flow
$\mathfrak{u}_{\mathbf{X}_{L}}$ will in general result in a flow
$\mathfrak{u}_{\mathbf{Y}}$ generated by a $\mathbf{Y}$ that is
\textit{not} a SOLVF. It need not even be a solution of the energy
equation. By necessity, general solutions of the energy equation must
be considered, leading us to consider the collection of solutions
\begin{equation}
\overline{\mathcal{S}\hbox{ol}}
:=\{\overline{\mathbf{X}}_{EL}\in\mathcal{S}\hbox{ol}  
\ \vert \ [\mathbf{G},
  \overline{\mathbf{X}}_{EL}]\in
\left[\mathbf{T}_{\mathfrak{u}}\mathbb{P}_L\right]^v \>\>
\>\> \forall \mathbf{G}\in
\mathcal{G}\}.  
\nonumber
\end{equation}
This collection generates the family of integral flows  
\begin{equation}
  \mathcal{O}_{EL}(\mathfrak{u}_0) := \bigg\{\mathfrak{u}(t) \ \bigg\vert \
  \frac{d\mathfrak{u}}{dt}=\overline{\mathbf{X}}_{EL}(\mathfrak{u}),
  \overline{\mathbf{X}}_{EL}\in\overline{\mathcal{S}\hbox{ol}},
  \>\>\hbox{and }
  \mathfrak{u}(t_0)=\mathfrak{u}_0\bigg\}.
  \nonumber
\end{equation}
Importantly, if $\mathbf{P}\in\mathcal{S}\hbox{ym}$, then 
\begin{equation}
  i_{[\mathbf{X}_E,
      \mathbf{P}]}\mathbf{\Omega}_L=i_{\mathbf{P}}\mathbf{d}\mathbf{\beta}[\mathbf{X}_E]=0.
  \nonumber
\end{equation}
As such, we find that 

\begin{lemma} \label{X_EL} $[\overline{\mathbf{X}}_{EL}, \mathbf{P}] \in
  \overline{\hbox{ker } \mathbf{\Omega}_L(\mathfrak{u})}$ for all
  $\mathbf{P}\in \mathcal{S}\hbox{ym}$.  
\end{lemma}

\noindent It then follows that

\begin{theorem}\label{Group-EL} $\hbox{Gr}_{\mathcal{S}\hbox{ym}}$ forms a group of
  symmetry transformations of $\mathcal{O}_{EL}(\mathfrak{u}_0)$.
\end{theorem}

\noindent Proof of both assertions can be found in \cite{ADS2020}.

The generators of the generalized Lie symmetry for
$\mathcal{O}_{EL}(\mathfrak{u}_0)$ are 
thus given by $\mathcal{S}\hbox{ym}$. The corresponding solutions
to the Euler-Lagrange equations that have this symmetry are given by
$\overline{\mathcal{S}\hbox{ol}}$, and a vector
$\overline{\mathbf{X}}_{EL} \in \overline{\mathcal{S}\hbox{ol}}$ is
called a \textbf{second-order, Euler-Lagrange vector field (SOELVF)}. It has
the general form,
\begin{equation}
  \overline{\mathbf{X}}_{EL}=\overline{\mathbf{X}}_{L} +
  \sum_{m=1}^{N_0} u^m(\mathfrak{u}) \left[\mathbf{P}_{(m)}\right],
  \label{EL}
\end{equation}
where $u^m(\mathfrak{u})\in\overline{\mathcal{F}}$ and
$\{[\mathbf{P}_{(n)}], n=1, \dots, N_0\}$ is a choice of basis 
for $\overline{\hbox{ker
  }\mathbf{\Omega}_L(\mathfrak{u})}/\mathcal{G}$. The vector field
$\overline{\mathbf{X}}_L$ is constructed from the second order
Lagrangian vector field $\mathbf{X}_L$ and vectors in $\hbox{ker
}\mathbf{\Omega}_L(\mathfrak{u})$ by requiring
$\overline{\mathbf{X}}_L\in\overline{\mathcal{S}\hbox{ol}}$. This
construction is described in \cite{ADS2020}; we will only need the
existence of such a vector field in this paper.

\section{Generalized Lie Symmetries of the Action and its Impact on
  Dynamics\label{&A-S}} 

We now turn our attention to the generators of the
generalized Lie symmetry of the action, and the impact this symmetry
has on the evolution of dynamical systems.

\subsection{The Generalized Lie Symmetry of the Action\label{SandA}}

In determining the conditions (as listed in \textbf{Lemma \ref{Action-Sym}}) under
which the action admits a generalized Lie symmetry, the understanding
that the action must have this symmetry for all possible paths on
$\mathbb{Q}$ played an essential role. By necessity, these
conditions could only be placed on $\rho_L$, and not on 
$\dot{\rho}_L$; unlike $\rho_L$, $\dot{\rho}_L$ depends explicitly on the
evolution of a particular path, while the symmetry must hold
for all paths. We note, however, that the family $\mathcal{O}_{EL}$ of
trajectories determined by the Euler-Lagrange equations of motion also consists of
paths on $\mathbb{Q}$, and as such the generalized Lie symmetry of the
action is also a symmetry of $\mathcal{O}_{EL}$. Importantly, how these
trajectories evolve with time is known, and as such, the $\dot{\rho}_L$ for a
given $\rho_L$ is also known for these trajectories. With this
understanding, and after comparing \textbf{Lemma \ref{Action-Sym}} and
the results of \textbf{Lemma \ref{AllS}} with \textbf{Lemma \ref{GS}},
we conclude that the generators of the generalized Lie symmetry of the
action must also be generators of the generalized Lie symmetry of the
Euler-Lagrange equations of motion. This leads us to consider the
following collection of vectors.  
\begin{equation}
  \mathcal{S}\hbox{ym}\mathcal{L} = \{\mathbf{P}\in \overline{\hbox{ker
    }\mathbf{\Omega}_L(\mathfrak{u})}/\mathcal{G}\ \vert\
  \gamma_{\mathbf{P}}^{[1]}=\langle \mathbf{\beta}\vert\mathbf{P}\rangle
    =0 \hbox{ on }\mathbb{P}_L \}.
    \nonumber
\end{equation}
We will also need  $N_{\mathcal{S}\hbox{ym}\mathcal{L}}=\hbox{dim
}(\mathcal{S}\hbox{ym}\mathcal{L})$ in the following. 

\begin{lemma}\label{subset}
  $\mathcal{S}\hbox{ym}\mathcal{L}\subset\mathcal{S}\hbox{ym}$.
  
  \begin{proof}
    Let $\{\mathbf{P}_{(l)}, l = 1, \dots, N_0 \}$ be a basis of
    $\overline{\hbox{ker
      }\mathbf{\Omega}_L(\mathfrak{u})}/\mathcal{G}$ such that
    $\mathbf{P}_{(l)}\in\mathcal{S}\hbox{ym}\mathcal{L}$ for
    $l= 1, \dots, N_{\mathcal{S}\hbox{ym}\mathcal{L}}$. We may choose the basis of
    $\mathcal{C}$ such that $\langle \mathbf{\Theta}^{(m)}_q\vert
    \mathbf{P}_{(l)}\rangle=\delta^{(m)}_{(l)}$. Then for any
    $\mathbf{P}_{(n)}\in\mathcal{S}\hbox{ym}$, we see from
    Eq.~$(\ref{beta})$ that, 
    \begin{equation}
      \langle\mathbf{d}\mathbf{\beta}\vert \mathbf{P}_{(n)}\otimes\mathbf{Y}\rangle= \sum_{m=1}^{N_0}
      \left(
      \langle\mathbf{d}\gamma_m^{[1]}\vert
      \mathbf{P}_{(n)}\rangle \langle \mathbf{\Theta}^{(m)}_q\vert
      \mathbf{Y}\rangle-
      \langle\mathbf{d}\gamma_m^{[1]}\vert\mathbf{Y}
      \rangle \langle \mathbf{\Theta}^{(m)}_q\vert
      \mathbf{P}_{(n)}\rangle+
      \gamma_m^{[1]}\langle\mathbf{d}\mathbf{\Theta}^{(m)}_q\vert
      \mathbf{P}_{(n)}\otimes\mathbf{Y}\rangle 
      \right),
      \nonumber
    \end{equation}
    for any $\mathbf{Y}\in \mathbf{T}\mathbb{P}_L$. The last term
    vanishes on the first-order constraint manifold
    $\mathbb{P}_L^{[1]}$, while for the second term, 
    $\langle\mathbf{d}\gamma_m^{[1]}\vert\mathbf{Y} \rangle
    \langle \mathbf{\Theta}^{(m)}_q\vert \mathbf{P}_{(n)}\rangle=
    \langle\mathbf{d}\gamma_n^{[1]}\vert\mathbf{Y} \rangle
    \delta ^{(m)}_{(n)}$. But as
    $\mathbf{P}_{(n)}\in\mathcal{S}\hbox{ym}\mathcal{L}$,
    $\gamma_n^{[1]}=0$ on $\mathbb{P}_L$, and this term
    vanishes as well. Finally, for the first term,
    $\langle\mathbf{d}\mathbf{\gamma}_m^{[1]}\vert
    \mathbf{P}_{(n)}\rangle= \mathbf{P}_{(n)}\mathbf{P}_{(m)}E=
    [\mathbf{P}_{(n)},\mathbf{P}_{(m)}]E+\mathbf{P}_{(m)}\mathbf{P}_{(n)}E$. But
    $\gamma_n^{[1]}=\mathbf{P}_{(n)}E=0$ on $\mathbb{P}_L$, while
    $\overline{\hbox{ker }\mathbf{\Omega}(\mathfrak{u})}$ is
    involutive. There then exists a $\mathbf{P}_{(nm)}\in
    \overline{\hbox{ker }\mathbf{\Omega}(\mathfrak{u})}$ such that
    $\mathbf{P}_{(nm)}=[\mathbf{P}_{(n)}, \mathbf{P}_{(m)}]$. As
    $\mathbf{P}_{(nm)}E:=\gamma_{(nm)}^{[1]}$, this
    $\gamma_{(nm)}^{[1]}$ must be a linear
    combination of first-order constraint functions, and they also vanish
    on $\mathbb{P}_L^{[1]}$. It then follows that
    $\langle\mathbf{d}\mathbf{\beta}\vert
    \mathbf{P}_{(n)}\otimes\mathbf{Y}\rangle=0$ on
    $\mathbb{P}_L^{[1]}$, and
    $\mathbf{P}_{(n)}\in\mathcal{S}\hbox{ym}$.
  \end{proof}
\end{lemma}

If $\mathbf{P}_1, \mathbf{P}_2 \in \mathcal{S}\hbox{ym}\mathcal{L}$, then
$\gamma^{[1]}_{[\mathbf{P}_1,\mathbf{P}_2]} =\mathbf{P}_1\mathbf{P}_2E-
\mathbf{P}_2\mathbf{P}_1E=\mathbf{P_1}\gamma_{\mathbf{P}_2}-
\mathbf{P_2}\gamma_{\mathbf{P}_1}=0$, and thus
$\mathcal{S}\hbox{ym}\mathcal{L}$ is involutive. Then for  
each $\mathbf{P}\in\mathcal{S}\hbox{ym}\mathcal{L}$ we once again have the
one-parameter subgroup
$\sigma^{\mathcal{S}\hbox{ym}\mathcal{L}}_{\mathbf{P}}(\epsilon,\mathfrak{u})$
define as the integral flow of
\begin{equation}
  \frac{\mathbf{d}\sigma^{\mathcal{S}\hbox{ym}\mathcal{L}}_{\mathbf{P}}}{\mathbf{d}\epsilon}
  :=\mathbf{P},
  \nonumber
\end{equation}
with
$\sigma^{\mathcal{S}\hbox{ym}\mathcal{L}}_{\mathbf{P}}(0,\mathfrak{u})=\mathfrak{u}$
for $\mathfrak{u}\in\mathbb{P}_L$. The collection of such subgroups
gives the Lie group $\hbox{Gr}_{\mathcal{S}{ym}\mathcal{L}}$. As
$\mathcal{S}{ym}\mathcal{L}\subset \mathcal{S}{ym}$,
$\hbox{Gr}_{\mathcal{S}{ym}\mathcal{L}}$ is a Lie subgroup of
$\hbox{Gr}_{\mathcal{S}{ym}}$. It then follows from \textbf{Theorem \ref{Group-EL}}
that $\hbox{Gr}_{\mathcal{S}\hbox{ym}\mathcal{L}}$ also forms a group of
  symmetry transformations of $\mathcal{O}_{EL}(\mathfrak{u}_0)$. As
  the family $\mathcal{O}_{EL}(\mathfrak{u}_0)$ of trajectories are
  paths on $\mathbb{Q}$, and as the symmetry transformation of the
  action must be the same for all paths on $\mathbb{Q}$, it also follows that, 

\begin{theorem}\label{Group-L}
  $\hbox{Gr}_{\mathcal{S}\hbox{ym}\mathcal{L}}$ forms the group of
  symmetry transformations of the action $S$.
\end{theorem}

\subsection{Symmetries and Dynamics\label{SandD}}

While $\mathcal{O}_{EL}(\mathfrak{u}_0)$ gives the family of integral
flows on which both $\hbox{Gr}_{\mathcal{S}\hbox{ym}}$ and
$\hbox{Gr}_{\mathcal{S}\hbox{ym}\mathcal{L}}$ act, a general flow in
$\mathcal{O}_{EL}(\mathfrak{u}_0)$ need not be confined to $\mathbb{P}_{L}^{\left[1\right]
}$, and yet this is  the submanifold on which the solutions 
$\overline{\mathbf{X}}_{EL}\in mathcal{S}{\mathrm{ol}}$ of the energy
equations exist. In such cases it is necessary to jointly choose a
SOELVF $\overline{\mathbf{X}}_{EL}$ and a submanifold 
of $\mathbb{P}_{L}^{\left[  1\right]  }$ on which the resultant flow
$\mathfrak{u}_{\overline{\mathbf{X}}_{EL}}$ will be confined. This is
done through the implementation of a constraint algorithm, one of
which was proposed in \cite{ADS2020}. In that paper the product of this
algorithm was the most that could be said about the general structure 
of SOELVFs that have integral flow fields which lie on
$\mathbb{P}_L^{[1]}$. Here, with the results obtained in
\textbf{Section \ref{SandA}}, we can say much more, and we will see
that the presence of a generalized Lie symmetry of the action
greatly restricts the structure of the SOELVFs that such systems can have.

Following \cite{ADS2020}, we introduce for a 
$\overline{\mathbf{X}}_{EL}\in\overline{\mathcal{S}\hbox{ol}}$ the
notation 
\begin{equation}
\overline{\mathbf{X}}_{EL}^{[1]} := \overline{\mathbf{X}}_{EL}, \>
\overline{\mathbf{X}}_{L}^{[1]} := \overline{\mathbf{X}}_{L}, \>
\mathbf{P}^{[1]}_{(n)} := \mathbf{P}_{(n)}, \> u_{[1]}^m := u^m, \>
N_0^{[1]} := N_0,
\nonumber
\end{equation}
when the constraint algorithm is implemented, with the superscript $[1]$ denoting
the first iteration of this algorithm. (This notation is only used in
this section.) In addition, we choose 
$\mathbf{P}^{[1]}_{(n)}\in \mathcal{S}\hbox{ym}\mathcal{L}$ for
$n=1,\dots, N_{\mathcal{S}\hbox{ym}\mathcal{L}}$.

For the integral flow field of $\overline{\mathbf{X}}_{EL}$ to lie on 
$\mathbb{P}_L^{[1]}$,
\begin{equation}
  \mathfrak{L}_{\overline{\mathbf{X}}_{EL}}\mathbf{\beta}=0,
  \label{stable}
\end{equation}
which reduces to
$\mathfrak{L}_{\overline{\mathbf{X}}_{EL}}\gamma_n^{[1]}=0$ on $\mathbb{P}_L^{[1]}$.
This is called the \textbf{constraint condition}. 
As both $u_{[1]}^n,
\gamma^{[1]}_n\in \overline{\mathcal{F}}$,
$\left[\mathbf{P}_{(n)}^{[1]}\right]\gamma^{[1]}_m = 
\mathbf{P}_{(n)}\gamma^{[1]}_m$, and after making use of the general
form of a SOELVF given in Eq.~$(\ref{EL})$, Eq.~$(\ref{stable})$ reduces to
\begin{equation}
  \sum_{m=1}^{N_0} \Gamma^{[1]}_{nm} u^m_{[1]} =
  -\left\langle \mathbf{d} \gamma^{[1]}_n\Big\vert
  \overline{\mathbf{X}}^{[1]}_{L}\right\rangle, \>\hbox{with }
  \Gamma^{[1]}_{nm} := \left\langle
  \mathbf{d}\gamma^{[1]}_n\Big\vert \mathbf{P}^{[1]}_{(m)}\right\rangle.
  \label{first-order}
\end{equation}
Since $\left\langle
\mathbf{d}\gamma^{[1]}_n\Big\vert
\mathbf{P}^{[1]}_{(m)}\right\rangle=\mathbf{P}^{[1]}_{(m)}\mathbf{P}^{[1]}_{(n)}E=
       [\mathbf{P}^{[1]}_{(m)}, \mathbf{P}^{[1]}_{(n)}]E
+\Gamma^{[1]}_{mn}$.
But $\overline{\hbox{ker }\mathbf{\Omega}_L(\mathfrak{u})}$ is
involutive, and thus $[\mathbf{P}^{[1]}_{(m)},
  \mathbf{P}^{[1]}_{(n)}]E$ is a linear combination of first-order
Lagrangian constraints. As these constraints vanishes on
$\mathbb{P}_L^{[1]}$, 
$\Gamma^{[1]}_{nm}=\Gamma^{[1]}_{mn}$ on the first-order constraint
manifold.

Next, when $n=1, \dots, N_{\mathcal{S}\hbox{ym}\mathcal{L}}$,
$\mathbf{P}_{(n)}^{[1]}\in\mathcal{S}\hbox{ym}\mathcal{L}$, and
$\gamma_n^{[1]}=0$. Thus, $\Gamma^{[1]}_{nm}=0$ when $n=1, \dots,
N_{\mathcal{S}\hbox{ym}\mathcal{L}}$, and as $\Gamma^{[1]}_{nm}$ is a
symmetric matrix on $\mathbb{P}_L^{[1]}$, $\Gamma^{[1]}_{mn}=0$ for
these values of $n$ as well. Thus while
$\Gamma^{[1]}_{nm}$ is a $N_0\times N_0$ matrix, 
the only nonzero components of this matrix lie in the
$\left(N_0-N_{\mathcal{S}\hbox{ym}\mathcal{L}}\right)\times
\left(N_0-N_{\mathcal{S}\hbox{ym}\mathcal{L}}\right)$ submatrix
$\bar{\Gamma}^{[1]}_{\overline{n}\,\overline{m}}:=
\left\langle\mathbf{d}\gamma^{[1]}_{\overline{n}+N_{\mathcal{S}\hbox{ym}\mathcal{L}}}\bigg\vert
  \mathbf{P}^{[1]}_{\overline{m}+N_{\mathcal{S}\hbox{ym}\mathcal{L}}}\right\rangle$
where $\overline{n}, \overline{m} = 1, \dots,
N_0-N_{\mathcal{S}\hbox{ym}\mathcal{L}}$. As $\left\langle \mathbf{d}
\gamma^{[1]}_n\Big\vert
\overline{\mathbf{X}}^{[1]}_{L}\right\rangle=0$ as well when $n=1,
\dots, N_{\mathcal{S}\hbox{ym}\mathcal{L}}$, Eq.~$(\ref{first-order})$
  reduces to  
\begin{equation}
  \sum_{\overline{m}=1}^{N_0-N_{\mathcal{S}\hbox{ym}\mathcal{L}}}
  \bar{\Gamma}^{[1]}_{\overline{n}\bar{m}}
  u^{\overline{m}+N_{\mathcal{S}\hbox{ym}\mathcal{L}}}_{[1]} = -\left\langle \mathbf{d}
  \gamma^{[1]}_{\overline{n}+N_{\mathcal{S}\hbox{ym}\mathcal{L}}}\Big\vert 
  \overline{\mathbf{X}}^{[1]}_{L}\right\rangle.
  \label{red-first-order}
\end{equation}
It is then readily apparent that the
$N_{\mathcal{S}\hbox{ym}\mathcal{L}}$ arbitrary functions $u^m_{[1]}$
for $m=1, \dots, N_{\mathcal{S}\hbox{ym}\mathcal{L}}$ are not
determined at this iteration, while $r^{[1]} =\hbox{rank }
\bar{\Gamma}^{[1]}_{\overline{n}\,\overline{m}}$ of the $u^m_{[1]}$ for
$m>N_{\mathcal{S}\hbox{ym}\mathcal{L}}$ are. There
are then $N_0^{[2]} :=N_0^{[1]}-r^{[1]}$
  \textbf{second-order Lagrangian constraint functions} 
\begin{equation}
  \gamma^{[2]}_{n_{[2]}} := \left\langle
  \mathbf{d}\gamma^{[1]}_{n_{[2]}}\Big\vert 
  \overline{\mathbf{X}}_{L}^{[1]}\right\rangle, n_{[2]}=1, \cdots,
  N_0^{[2]},
  \nonumber
\end{equation}
with the conditions $\gamma^{[2]}_{n_{[2]}}=0$ imposed if necessary. In general
there will be $I_{[2]}:=  
\hbox{rank }\left\{\mathbf{d}\gamma^{[1]}_{n_{[1]}},
  \mathbf{d}\gamma^{[2]}_{n_{[2]}}\right\}$
independent functions in $\hbox{C}^{[2]}_L :=
\hbox{C}^{[1]}\cup \left\{\gamma^{[2]}_{n_{[2]}}\ \vert
\ n_{[2]} = 1, \dots, N_0^{[2]}\right\}$, and $\mathbb{P}_L^{[1]}$ is
reduced to the \textbf{second-order constraint submanifold}, 
\begin{equation}
  \mathbb{P}_L^{[2]} := \left\{\mathfrak{u}\in\mathbb{P}_L^{[1]}
  \ \Big\vert \ \gamma^{[2]}_{[n_2]}(\mathfrak{u})=0, n_{[2]} = 1, \dots,
  N_0^{[2]} \right\}, 
  \nonumber
\end{equation}
where dim $\mathbb{P}^{[2]}_L = 2D-I_{[2]}$. At this point, there are two
possibilities. If $I_{[2]}=I_{[1]}$ or
$I_{[2]}=2D$, the iterative process stops, and no new Lagrangian
constraints are introduced. If not, the process continues. 

For the second iteration in the constraint algorithm, we choose a basis
$\left\{\mathbf{P}_{(n)}^{[2]}\right\}$ for $\overline{\hbox{ker
  }\mathbf{\Omega}_L(\mathfrak{u})}/\mathcal{G}$ and the arbitrary functions
$\left\{u_{[2]}^m\right\}$ such that for $m=1, \dots, N_0^{[2]}$,
$u_{[2]}^m$ are linear combinations of $u_{[1]}^m$ that lie in the kernel
$\Gamma^{[1]}_{nm}$. We once again require that
$\mathbf{P}_{(n)}^{[2]}\in\mathcal{S}\hbox{ym}\mathcal{L}$
for $n=1, \dots, N_{\mathcal{S}\hbox{ym}\mathcal{L}}$.  
Then
\begin{equation}
\overline{\mathbf{X}}_{EL}^{[2]} = \overline{\mathbf{X}}_{L}^{[2]} +
\sum_{m=1}^{N_0^{[2]}} u_{[2]}^m \left[\mathbf{P}_{(m)}^{[2]}\right],
\nonumber
\end{equation}
with
\begin{equation}
  \overline{\mathbf{X}}_{L}^{[2]} = \overline{\mathbf{X}}_{L}^{[1]} +
  \sum_{m=N_0^{[2]}+1} ^{N_0^{[1]}}u^m_{[2]}\left[\mathbf{P}^{[2]}_{(m)}\right].
  \nonumber
\end{equation}
Here, the functions $u^m_{[2]}$ for $m = N_0^{[2]}+1, \dots, N_0^{[1]}$ have been
determined through the constraint analysis of $\gamma^{[1]}_n$. 

As shown in \cite{ADS2020}, $\mathbf{G}u_{[1]}^m=0$. Similarly,
  $\mathbf{G}\gamma^{[2]}_n = \mathfrak{L}_{[\mathbf{G},
      \overline{\mathbf{X}}_{EL}]}\mathbf{d}\gamma^{[2]}_n=0$. Clearly
  $\gamma^{[2]}_n\in\overline{\mathcal{F}}$ and we may require
  $u^m_{[2]}\in\overline{\mathcal{F}}$ as well. It 
  then follows that $\left[\mathbf{P}^{[2]}_{(n)}\right]\gamma^{[2]}_m 
=\mathbf{P}^{[2]}_{(n)}\gamma^{[2]}_m$, and imposing Eq.~$(\ref{stable})$
on $\gamma^{[2]}_n $, gives
\begin{equation}
  \sum_{m=1}^{N_0^{[2]}}\Gamma^{[2]}_{nm} u^m_{[2]} =
  -\left\langle \mathbf{d}\gamma^{[2]}_n\Big\vert
  \overline{\mathbf{X}}^{[2]}_{L}\right\rangle,\>\hbox{where }
  \Gamma^{[2]}_{nm} := \left\langle
  \mathbf{d}\gamma^{[2]}_n\Big\vert \mathbf{P}^{[2]}_{(m)}\right\rangle,
\>\> n=1, \dots,
N_0^{[2]}.
\label{second-order}
\end{equation}
Once again, $\Gamma^{[2]}_{nm}=\Gamma^{[2]}_{mn}$, but now on the constraint
manifold $\mathbb{P}^{[2]}_L$. Moreover, since
$\gamma^{[2]}_n=\gamma^{[1]}_n=0$ for $n=1, \dots,
N_{\mathcal{S}\hbox{ym}\mathcal{L}}$,  
$\Gamma^{[2]}_{nm}=0=\Gamma^{[2]}_{mn}$, and $\left\langle
\mathbf{d}\gamma_n^{[2]}\Big\vert\overline{\mathbf{X}}^{[2]}_L\right\rangle
=0$. There is once again a reduction of Eq.~$(\ref{second-order})$,
and we are left with 
\begin{equation}
  \sum_{\overline{m}=1}^{N_0^{[2]}-N_{\mathcal{S}\hbox{ym}\mathcal{L}}}
    \overline{\Gamma}^{[2]}_{\overline{n}\,\overline{m}} u^{\overline{m}+N_{\mathcal{S}\hbox{ym}\mathcal{L}}}_{[2]} =
    -
    \langle \mathbf{d}\gamma^{[2]}_{\overline{n}+N_{\mathcal{S}\hbox{ym}\mathcal{L}}}\vert
  \overline{\mathbf{X}}^{[2]}_{L} \rangle.
\nonumber
\end{equation}
where
$\bar{\Gamma}^{[2]}_{\overline{n}\,\overline{m}}:=
\left\langle\mathbf{d}\gamma^{[2]}_{\overline{n}+N_{\mathcal{S}\hbox{ym}\mathcal{L}}}
\bigg\vert\mathbf{P}^{[2]}_{(\overline{m}+N_{\mathcal{S}\hbox{ym}\mathcal{L}})}\right\rangle$. As
before, the $N_{\mathcal{S}\hbox{ym}\mathcal{L}}$ arbitrary functions
$u^{m}_{[2]}$ are not determined, while $r^{[2]} := \hbox{rank
}\bar{\Gamma}^{[2]}_{\overline{n}\,\overline{m}}$ of the
remaining $u^m_{[2]}$ for $m>N_{\mathcal{S}\hbox{ym}\mathcal{L}}$
are. There are now 
$N^{[3]}_0=N_0^{[2]}-r^{[2]}$ \textbf{third-order Lagrangian
  constraint functions},
\begin{equation}
  \gamma^{[3]}_{n_{[3]}} =
  \left\langle\mathbf{d}\gamma^{[2]}_{n_{[3]}}\Big\vert
  \overline{\mathbf{X}}^{[2]}_{L}\right\rangle, \> n_{[3]} = 1, \dots,
  N_0^{[3]},
  \nonumber
\end{equation}
with the conditions $\gamma^{[3]}_{n_{[3]}}=0$ 
imposed if necessary. With 
\begin{equation}
I_{[3]} := \hbox{rank } \left\{ 
  \mathbf{d}\gamma_{n_{[1]}}^{[1]},
 \mathbf{d}\gamma_{n_{[2]}}^{[2]},
 \mathbf{d}\gamma_{n_{[3]}}^{[3]}
 \right\},
 \nonumber
\end{equation}
independent functions in $\hbox{C}^{[3]}_L :=
\hbox{C}_L^{[2]}\cup\left\{\gamma_{n_{[3]}}^{[3]}, n_{[3]}=1,
\dots, N_0^{[3]}\right\}$, we now have the \textbf{third-order
  constraint submanifold},  
\begin{equation}
  \mathbb{P}_L^{[3]}:=\left\{\mathfrak{u}\in \mathbb{P}^{[2]}_L
  \ \Big\vert \ \gamma^{[3]}_{n_{[3]}}(\mathfrak{u})=0, n_{[3]}=1,
  \dots, N_0^{[3]}\right\}.
  \nonumber
\end{equation}
Once again, the process stops when
$I_{[3]}=I_{[2]}$ or $I_{[3]}=2D$. However, if $I_{[2]}<I_{[3]}<2D$, the
process continues until at the $n_F$-iteration when either
$I_{[n_F]}=I_{[n_F]-1}$ or $I_{[n_F]}=2D$. 

Following \cite{ADS2020}, the end result of this algorithm is 
\begin{enumerate}
\item{A submanifold $\mathbb{P}^{[n_F]}_L\subset \mathbb{P}_L$ on which
  dynamics takes place.} 
  \item{A collection
    $\hbox{C}_L^{[n_F]}\subset\overline{\mathcal{F}}$ of constraint
    functions of order $1$ to $n_F$.} 
\item{A second-order, Euler-Lagrange vector field
  \begin{equation}
    \overline{\mathbf{X}}_{EL}^{[n_F]}=\overline{\mathbf{X}}_{L}^{[n_F]} +
    \sum_{m=1}^{N^{[n_F]}_0}u^m_{[n_F]}(\mathfrak{u})\left[\mathbf{P}_{(m)}^{[n_F]}\right],  
    \nonumber
  \end{equation}
  with $N_0^{[n_F]}\ge N_{\mathcal{S}\hbox{ym}\mathcal{L}}$ arbitrary functions
  $u^m_{[n_F]}(\mathfrak{u})\in\overline{\mathcal{F}}$ for $m=1,
  \dots, N_0^{[n_F]}$, and 
  \begin{equation}
    \overline{\mathbf{X}}_{L}^{[n_F]} =
    \overline{\mathbf{X}}_{L}^{[1]}+
    \sum_{m=N_0^{[n_F]}+1}^{N_0^{[1]}}u^m_{[n_F]}(\mathfrak{u})
    \left[\mathbf{P}_{(m)}^{[n_F]}\right],  
    \nonumber
  \end{equation}
  where the $N_0^{[1]}-N_0^{[n_F]}$ functions
  $u^m_{[n_F]}(\mathfrak{u})\in\overline{\mathcal{F}}$, 
  $m=N_0^{[n_F]}+1, \dots, N_0^{[1]}$, have been uniquely
  determined through the constraint algorithm.} 
\end{enumerate}
We assume that the rank of $\Gamma^{[l]}_{nm}$ is constant on
$\mathbb{P}_L$ for each $l=1, \dots, n_F$, and that
$\mathbb{P}^{[n_F]}_L$ is non-empty. 

The end result of the constraint algorithm
$\overline{\mathbf{X}}_{EL}^{[n_F]}$ is still a SOELVF, and we define
the collection of such vector fields as
\begin{equation}
  \overline{\mathcal{S}\hbox{ol}}_{\mathbb{P}^{[n_F]}_L} :=
  \{\overline{\mathbf{X}}_{EL}\in\overline{\mathcal{S}\hbox{ol}} \>\>\vert\>
  \mathfrak{L}_{\overline{\mathbf{X}}_{EL}}\mathbf{\beta}=0\}.
  \nonumber
\end{equation}
Importantly, $\hbox{dim }\overline{\mathcal{S}\hbox{ol}}_{\mathbb{P}^{[n_F]}_L}\ge
N_{\mathcal{S}\hbox{ym}\mathcal{L}}$. 

\section{The Generalized Lie Symmetries of Three Dynamical
  Systems\label{&Exam}}

Three examples of dynamical systems with almost regular Lagrangians
were introduced in \cite{ADS2020}. In that paper the focus of these
examples was on the explicit construction of the dynamical structures 
needed to describe and predict motion in the Lagrangian phase space,
and to show that these structures are projectable to 
the Hamiltonian phase space. We return to these examples here, but
with the focus now being on the generalized Lie symmetries of each, and the
application of the results we have found in this paper. In particular,
we are in interested in the dimensionality of the symmetry groups for
each of the systems as compared to the dimensionality of
$\overline{\mathcal{S}\hbox{ol}}_{\mathbb{P}^{[n_F]}_L}$ of each. A
summary of our results can be found in Table $\ref{Table}$

\subsection{A Lagrangian With and Without a Generalized Lie Symmetry\label{&special}}  

Whether the action
\begin{equation}
  S_{1}:=\int\left[\frac{1}{2}m\left(\frac{d\widehat{q}}{dt}\right)^2
    -V(q^a)\right]dt, 
  \nonumber
\end{equation}
with $\vert q\vert=\sqrt{q^a q_a}$ and $\widehat{q}^a := q^a/\vert
q\vert$, $a=1, \dots, D$, has a generalized Lie symmetry depends on the
choice of potential $V(q)$. With one choice both the Lagrangian and
the Euler-Lagrange equations of 
motion have a generalized gauge symmetry; with a second choice the equations
of motion has a generalized Lie symmetry while the Lagrangian
does not; and with a third choice neither the action nor the equations of motion have
a symmetry. Irrespective of the choice of $V(q)$, however, $L$ is
singular, demonstrating that while all actions with a 
generalized Lie symmetry have a singular Lagrangian, not all singular
Lagrangians have a generalized Lie symmetry. 

Defining $\Pi_{ab}(q):= \delta_{ab} - \widehat{q}_a\widehat{q}_b$, we find
\begin{equation}
  \mathbf{\Omega}_M = \frac{m}{\vert q\vert^2}\Pi_{ab}(q)
  \mathbf{d}q^a\wedge\mathbf{d}v^b,\qquad
  \mathbf{\Omega}_F=\frac{m}{\vert q\vert^3}
  \left(\widehat{q}\cdot\mathbf{d}q\right)\wedge
  \left(v\cdot\Pi(q)\cdot\mathbf{d}q\right).
\nonumber\end{equation}
Then $\mathcal{C}$ and $\mathcal{G}$ are spanned by $\mathbf{U}^q_{(1)}
= \widehat{q}\cdot\mathbf{\partial}/\mathbf{\partial} q$ and  
  $\mathbf{U}^v_{(1)} =
  \widehat{q}\cdot\mathbf{\partial}/\mathbf{\partial} v$, 
respectively, while $\overline{\hbox{ker
  }\mathbf{\Omega}_L(\mathfrak{u})}$ is spanned by
$\mathbf{U}^v_{(1)}$ and  
\begin{equation}
  P_{(1)}=\widehat{q}\cdot\frac{\mathbf{\partial}\>\>\>}{\mathbf{\partial} q} +
    \frac{1}{\vert q\vert}v\cdot\frac{\mathbf{\partial}\>\>\>}{\mathbf{\partial}
      v}.
\nonumber\end{equation}
That $\hbox{dim }(\overline{\hbox{ker
  }\mathbf{\Omega}_L(\mathfrak{u})}/\mathcal{G})=1$ then follows.

The energy is
\begin{equation}
  E=\frac{1}{2}\frac{m}{\vert q\vert^2}v\cdot\Pi(q)\cdot v + V(q),
  \nonumber
\end{equation}
and there is only one first-order Lagrangian constraint,
\begin{equation}
  \gamma^{[1]}=\mathbf{U}^q_{(1)} V,
  \label{g-con}
\end{equation}
so that $\mathbf{\beta}[\mathbf{X}_{EL}] =
\gamma^{[1]}\mathbf{\Theta}^{(1)}_q$, where $\mathbf{\Theta}^{(1)}_q =
\widehat{q}\cdot \mathbf{d} q$. Using Eq.~$(\ref{g-con})$,
\begin{equation}
  \mathfrak{L}_{\mathbf{P}_{(1)}} \mathbf{\beta} =
  \mathbf{d}\left[\mathbf{U}^q_{(1)}V
    \right]-\frac{1}{\vert
    q\vert^2}\widehat{q}\cdot\frac{\partial\>\>\>}{\partial 
    q}\left(\Pi_a^{\>\>b}(q)\frac{\partial V}{\partial\widehat{q}^b}\right)
  \mathbf{d}q^a.
  \label{e84}
\end{equation}
Whether or not $\mathcal{S}\hbox{ym}$ or
$\mathcal{S}\hbox{ym}\mathcal{L}$ is empty therefore depends on the
symmetries of $V(q)$, as we would expect.  

It was found in \cite{ADS2020} that
\begin{equation}
  \overline{\mathbf{X}}_{L} = v\cdot{\Pi(q)}\cdot
  \frac{\mathbf{\partial}\>\>\>}{\mathbf{\partial} q} +
  \frac{(\widehat{q}\cdot v)}{\vert q\vert}
  v\cdot\frac{\mathbf{\partial}\>\>\>}{\partial v} -
  \frac{\vert q\vert^2}{m}\frac{\partial V}{\partial
    q}\cdot\Pi(q)\cdot\frac{\mathbf{\partial}\>\>\>}{\mathbf{\partial}v},
\nonumber\end{equation}
and a general SOELVF is given by
$\overline{\mathbf{X}}_{EL}=\overline{\mathbf{X}}_{L} + u(\mathfrak{u})
\left[\mathbf{P}_{(1)}\right]$, where
$u(\mathfrak{u})\in\overline{\mathcal{F}}$.
As the constraint 
algorithm gives 
\begin{equation}
  \mathfrak{L}_{\overline{\mathbf{X}}_{EL}}\gamma^{[1]} = v\cdot\Pi
  \cdot\frac{\partial \gamma^{[1]}}{\partial q} + u(\mathfrak{u})
  \mathbf{U}^q_{(1)}\gamma^{[1]},
  \label{stab}
\end{equation}
whether or not $u(\mathfrak{u})$ (which in turn determines the
dimensionality of
$\overline{\mathcal{S}\hbox{ol}}_{\mathbb{P}_L^{[n_f]}}$) is determined
by the constraint condition also depends on the symmetries of $V(q)$. 

There are three cases to consider.

\bigskip
\noindent{\textit{The symmetric potential}}
\bigskip
    
For $\mathbf{P}_{(1)}$ to generate a generalized Lie symmetry of the
Euler-Lagrange equations of motion, 
\begin{equation}
  0=\frac{1}{\vert q\vert^2} \widehat{q}\cdot\frac{\partial \>\>\>}{\partial q} \left(\Pi_a^{\>\>b}(q)\frac{\partial
    V}{\partial \widehat{q}^b}\right),
\nonumber\end{equation}
and as such the potential must satisfy
\begin{equation}
  \frac{\partial V}{\partial\widehat{q}^a} = \frac{\partial
    V_{AS}(\widehat{q}^a)}{\partial \widehat{q}^a},
\nonumber\end{equation}
where $V_{AS}$ is a function of $\widehat{q}^a$ only. It follows that
$\mathbf{P}_{(1)}$ generates a generalized Lie symmetry iff $V(q^a) =
V_{Sph}(\vert q\vert)+V_{AS}(\widehat{q}^a)$, where $V_{Sph}$ is a function of
$\vert q\vert$ only. For this potential, $\mathcal{S}\hbox{ym}$ is 
one-dimensional, and is spanned by $\mathbf{P}_{(1)}$. 

The constraint condition Eq.~$(\ref{stab})$ for this potential
reduces to 
\begin{equation}
  0=u(\mathfrak{u})\frac{d^2V_{Sph}(q)}{d\vert q\vert^2},
\nonumber\end{equation}
which must be satisfied on $\mathbb{P}_L^{[1]}$. There are two possibilities.

\bigskip
\textit{Case 1:} $\frac{d^2V_{Sph}}{d\vert q\vert^2}=0$.
\bigskip

\noindent Then $V_{Sph}(\vert q\vert) = a\vert q\vert+b$, but since 
\begin{equation}
  \gamma^{[1]}=\frac{dV_{Sph}}{d\vert q\vert} =a,
\nonumber\end{equation}
the condition $\gamma^{[1]}=0$ requires $a=0$. It then follows that
$\gamma^{[1]} =0$ on $\mathbb{P}_L$, and thus
$\mathcal{S}\hbox{ym}\mathcal{L}$ is one-dimensional; it also is
spanned by $\mathbf{P}_{(1)}$. The potential is then $V(q) =
b+V_{AS}(\widehat{q}^a)$, and the Lagrangian is   
invariant under the transformation $q^a\to \alpha q^a$,
where $\alpha$ is an arbitrary, nonvanishing function on
$\mathbb{P}_L$. This Lagrangian therefore has a local conformal symmetry.
Importantly, the function $u(\mathfrak{u})$ is
not determined, and thus the dynamics of the 
particle is given only up to an arbitrary function. Then  
$\hbox{dim
}(\overline{\mathcal{S}\hbox{ol}}_{\mathbb{P}^{[n_F]}_L})=1$ as well,
and is also spanned by $\mathbf{P}_{(1)}$.

\bigskip
\textit{Case 2:} $\frac{d^2V_{Sph}}{d\vert q\vert^2}\ne0$.
\bigskip

\noindent In this case $u(\mathfrak{u})=0$, and the dynamics of the
particle is completely determined by its initial data;
$\overline{\mathcal{S}\hbox{ol}}_{\mathbb{P}^{[n_F]}_L}=\{\overline{\mathbf{X}}_{L}\}$.
The first-order Lagrangian constraint $\gamma^{[1]}$ does not vanish
automatically, but instead defines a surface
on $\mathbf{P}_L$, and it follows that
$\mathcal{S}\hbox{ym}\mathcal{L}=\emptyset$. Indeed, the action's lack
of a local gauge symmetry in this case can be seen explicitly.

Equation $(\ref{g-con})$ reduces to
\begin{equation}
  0=\widehat{q}\cdot\frac{\mathbf{\partial}V_{sph}}{\mathbf{\partial}q},
\nonumber\end{equation}
and for dynamics to be possible the set of solutions 
\begin{equation}
    \left\{R_i\in\mathbb{R} \ \Bigg\vert
    \ \frac{dV_{Sph}}{d\vert q\vert}\Bigg\vert_{R_i} =0\right\},
\nonumber\end{equation}
must be non-empty. Dynamics are on the surfaces $\vert q\vert
-R_i =0$ where the potential reduces to $V(q)=V_{Sph}(R_i) +
V_{AS}(\widehat{q}^a)$. This reduced potential has the same symmetry
as the potential $V_{AS}(\widehat{q}^a)$ in \textit{Case 1}, and it is for
this reason that the Euler-Lagrange equations of motion have the same
generalized Lie symmetry for the two cases. This is explicitly shown
in the appendix.

In \textit{Case 1} the action has a local conformal symmetry, while in
\textit{Case 2} it does not. (In \cite{ADS2020} it was
erroneously stated that in this case the action has a global rotational
symmetry.) The Lagrangian for the two cases do not have the same 
invariances, resulting in one case dynamics that are determined
only up to 
an arbitrary $u(\mathfrak{u})$, and in the other case to a
$u(\mathfrak{u})=0$ and dynamics
that are instead completely determined by the choice of initial data. 

\bigskip
\noindent{\textit{The asymmetric potential}}
\bigskip

For a general $V$ the second term in
Eq.~$(\ref{e84})$ does not vanish, $\mathbf{P}_{(1)}$ does not
generate a symmetry of the equations of motion, and 
$\mathcal{S}\hbox{ym}=\{\emptyset\}$. As before, $\gamma^{[1]}$ does
not vanish, and thus $\mathcal{S}\hbox{ym}\mathcal{L}=\{\emptyset\}$
as well. Furthermore, as Eq.$(\ref{stab})$ results in
\begin{equation}
  \overline{\mathbf{X}}_E =\overline{\mathbf{X}}_L-\frac{v\cdot\Pi
  \cdot\frac{\partial \gamma^{[1]}}{\partial q}}
  {q^2\mathbf{U}^q_{(1)}\gamma^{[1]}}[\mathbf{P}_{(1)}],
\nonumber\end{equation}
the dynamics of the particle is uniquely determined by its initial
data, and
$\overline{\mathcal{S}\hbox{ol}}_{\mathbb{P}^{[n_F]}_L}=\{\overline{X}_{EL}\}$
once again consists of a single point.

\subsection{A Lagrangian with Local
  Conformal Symmetry}

The action,
\begin{equation}
  S_{2} := \int\Bigg\{\frac{1}{2}m
  \left(\frac{d\widehat{q}_1}{dt}\right)^2+\frac{1}{2}m
  \left(\frac{d\widehat{q}_2}{dt}\right)^2+
  \frac{\lambda}{2}\left[\frac{q_1^a}{\vert q_2\vert}
  \frac{d\>\>}{dt}\left(\frac{q_{2a}}{\vert q_1\vert}\right) -
  \frac{q_2^a}{\vert q_1\vert}
  \frac{d\>\>}{dt}\left(\frac{q_{1a}}{\vert q_2\vert}\right)
  \right]
  \Bigg\} dt,
\nonumber\end{equation}
where $a=1, \dots, d$, $D=2d$, describes an interacting, two particle
system that is invariant under the local conformal transformation
$q_1^a \to \alpha(\mathfrak{u}) q^a_1$ and $q_2^a \to
\alpha(\mathfrak{u}) q^a_2$. 

With
\begin{widetext}
\begin{eqnarray}
  \mathbf{\Omega}_M &=&  \frac{m}{\vert q_1\vert^2} \Pi_{ab}(q_1)
  \mathbf{d}q_1^a\wedge \mathbf{d} v_1^b + \frac{m}{\vert q_2\vert^2}
  \Pi_{ab}(q_2) \mathbf{d}q_2^a\wedge \mathbf{d} v_2^b, \hbox{  and}
  \nonumber
  \\
  \mathbf{\Omega}_F &=& \frac{m}{\vert q_1\vert^3}
  \left(\widehat{q}_1\cdot\mathbf{d}
  q_1\right)\wedge\left(v_1\cdot\Pi(q_1)\cdot \mathbf{d} q_1\right) + 
  \frac{m}{\vert q_2\vert^3} \left(\widehat{q}_2\cdot\mathbf{d} 
  q_2\right)\wedge\left(v_2\cdot\Pi(q_2)\cdot \mathbf{d} q_2\right)- 
  \nonumber
  \\
  &{}&
  \frac{\lambda}{\vert q_1\vert \vert
    q_2\vert}\left[\mathbf{d}q_1^a\wedge\left(\Pi(q_2)\cdot 
  \mathbf{d} q_2\right)_a+ \left(\Pi(q_1)\cdot
  \mathbf{d} q_1\right)_a\wedge \mathbf{d}q_2^a -\left(\Pi(q_1)\cdot
  \mathbf{d} q_1\right)^a\wedge \left(\Pi(q_2)\cdot
  \mathbf{d} q_2\right)_a\right]-
  \nonumber
  \\
  &{}& \frac{\lambda}{\vert q_1\vert^2}
  \left(\widehat{q}_1\cdot\mathbf{d}q_1\right)\wedge
  \left(\widehat{q}_2\cdot\Pi(q_1) \cdot \mathbf{d}q_1\right) +
  \frac{\lambda}{\vert q_2\vert^2} 
  \left(\widehat{q}_2\cdot\mathbf{d}q_2\right)\wedge
  \left(\widehat{q}_1\cdot\Pi(q_2) \cdot \mathbf{d}q_2\right),
  \nonumber
\end{eqnarray}
\end{widetext}
$\mathcal{C}$ and $\mathcal{G}$ are two-dimensional, and
are spanned by
\begin{equation}
  \mathbf{U}^q_{(1)} =
  \widehat{q}_1\cdot\frac{\mathbf{\partial}\>\>\>}{\mathbf{\partial} q_1},
    \quad
  \mathbf{U}^q_{(2)} =
  \widehat{q}_2\cdot\frac{\mathbf{\partial}\>\>\>}{\mathbf{\partial} q_2},
  \quad \hbox{and}\quad 
  \mathbf{U}^v_{(1)} =
  \widehat{q}_1\cdot\frac{\mathbf{\partial}\>\>\>}{\mathbf{\partial} v_1},
    \quad
  \mathbf{U}^v_{(2)} =
  \widehat{q}_2\cdot\frac{\mathbf{\partial}\>\>\>}{\mathbf{\partial} v_2},
\nonumber\end{equation}
respectively. The reduced $\bar{F}=0$, and $\overline{\hbox{ker
  }\mathbf{\Omega}_L(\mathfrak{u})}$ is spanned by $\mathbf{U}^v_{(1)},
\mathbf{U}^v_{(2)}$, 
\begin{eqnarray}
  \mathbf{P}_{(+)} &=&
  q_1\cdot\frac{\mathbf{\partial}\>\>\>}{\mathbf{\partial} q_1}+q_2\cdot\frac{\mathbf{\partial}\>\>\>}{\mathbf{\partial} q_2}
  +
  v_1\cdot\frac{\mathbf{\partial}\>\>\>}{\mathbf{\partial}v_1} 
  +
  v_2\cdot\frac{\mathbf{\partial}\>\>\>}{\mathbf{\partial}v_2},
  \nonumber
\end{eqnarray}
and
\begin{eqnarray}
  \mathbf{P}_{(-)} &=&
  q_1\cdot\frac{\mathbf{\partial}\>\>\>}{\mathbf{\partial} q_1}-q_2\cdot\frac{\mathbf{\partial}\>\>\>}{\mathbf{\partial} q_2}
  +
  v_1\cdot\frac{\mathbf{\partial}\>\>\>}{\mathbf{\partial}v_1} 
  -
  v_2\cdot\frac{\mathbf{\partial}\>\>\>}{\mathbf{\partial}v_2} 
  -
  \nonumber
  \\
  &{}&
  2\frac{\lambda}{m}
  \left[\frac{\vert q_1\vert}{\vert
      q_2\vert}q_2\cdot\frac{\mathbf{\partial}\>\>\>}{\mathbf{\partial} 
      v_1}
    +\frac{\vert q_2\vert}{\vert q_1\vert}q_1\cdot
    \frac{\mathbf{\partial}\>\>\>}{\mathbf{\partial} v_2}\right].
  \nonumber
\end{eqnarray}
As such, $\hbox{dim }\overline{(\hbox{ker }\mathbf{\Omega}_L)}/\mathcal{G}=2$. 

The energy is
\begin{equation}
  E=\frac{1}{2}\frac{m}{\vert q_1\vert^2}v_1\cdot\Pi(q_1)\cdot v_1 +
  \frac{1}{2}\frac{m}{\vert q_2\vert^2}v_2\cdot\Pi(q_2)\cdot v_2.
\nonumber
\end{equation}
We find that $\gamma^{[1]}_{(+)} =0$ while
\begin{equation}
  \gamma^{[1]}_{(-)} =-\frac{2\lambda}{\vert q_1\vert \vert q_2\vert}\left(
  q_2\cdot \Pi(q_1)\cdot v_1 +
  q_1\cdot \Pi(q_2) \cdot v_2\right),
\nonumber\end{equation}
giving,
\begin{eqnarray}
  \mathbf{\beta}[\mathbf{X}_{EL}] &=& \frac{1}{2}\gamma^{[1]}_{(-)}
  \left(\frac{\mathbf{\Theta}_q^{(1)}}{\vert q_1\vert} -
  \frac{\mathbf{\Theta}_q^{(2)}}{\vert q_2\vert}\right).
  \nonumber
\end{eqnarray}
Then $\mathcal{S}\hbox{ym}\mathcal{L}$ is
one-dimensional and spanned by $\mathbf{P}_{(+)}$. 
As expected, $\mathfrak{L}_{\mathbf{P}_{(+)}}
\mathbf{\beta} = 0$. Because
\begin{equation}
  \mathfrak{L}_{\mathbf{P}_{(-)}} \mathbf{\beta} =
  -\frac{4\lambda}{m}\left[1-(\widehat{q}_1\cdot\widehat{q}_2)^2\right]\left(\frac{\mathbf{\Theta}_q^{(1)}}{\vert q_1\vert} -
  \frac{\mathbf{\Theta}_q^{(2)}}{\vert q_2\vert}\right),
  \nonumber
\end{equation}
$\mathcal{S}\hbox{ym}$ is also one-dimensional, and
is also spanned by $\mathbf{P}_{(+)}$. 

A general SOELVF is 
\begin{equation}
  \overline{\mathbf{X}}_{EL} = \overline{\mathbf{X}}_{L} -
  \frac{m}{8\lambda^2}\frac{\overline{\mathbf{X}}_{L}\gamma^{[1]}_{(-)}}{ 
    \left[1-(\widehat{q}_1\cdot\widehat{q}_2)\right]}\left[\mathbf{P}_{(-)}\right] +
  u^{(+)}(\mathfrak{u}) \left[\mathbf{P}_{(+)}\right],
  \label{103}
\end{equation}
where $u^{(+)}(\mathfrak{u})\in\overline{\mathcal{F}}$, and from \cite{ADS2020},
\begin{eqnarray}
\overline{\mathbf{X}}_{L} &=&
v_1\cdot\Pi(q_1)\cdot\frac{\mathbf{\partial}\>\>\>}{\mathbf{\partial}
  q_1} + v_2\cdot\Pi(q_2)\cdot\frac{\mathbf{\partial}\>\>\>}{\mathbf{\partial}
  q_2} +
\nonumber
\\
&{}&
\left(\frac{\widehat{q}_1\cdot v_1}{\vert q_1\vert}\right)
v_1\cdot\Pi(q_1)\cdot\frac{\mathbf{\partial}\>\>\>}{\mathbf{\partial} v_1} +
\left(\frac{\widehat{q}_2\cdot v_2}{\vert q_2\vert }\right)
v_2\cdot\Pi(q_2)\cdot\frac{\mathbf{\partial}\>\>\>}{\mathbf{\partial} v_2} 
+
\nonumber
\\
&{}&
\frac{\lambda}{m} \left(\frac{\vert q_1\vert}{\vert q_2\vert}
  v_2\cdot\Pi(q_2)\cdot\Pi(q_1)\cdot \frac{\mathbf{\partial}
    \>\>\>}{\mathbf{\partial} v_1}-\frac{\vert q_2\vert}{\vert q_1\vert}
  v_1\cdot\Pi(q_1)\cdot\Pi(q_2)\cdot \frac{\mathbf{\partial}
    \>\>\>}{\mathbf{\partial} v_2} \right),
  \nonumber
\end{eqnarray}
after the constraint algorithm is applied. Equation $(\ref{103})$ is
a consequence of the identity
$\langle\mathbf{d}\gamma^{[1]}_{(+)}\vert \overline{\mathbf{X}}_{L}\rangle=0$ and 
\begin{eqnarray}
 -\frac{1}{2\lambda} \langle\mathbf{d}\gamma^{[1]}_{(-)}\vert \overline{\mathbf{X}}_{L}\rangle&=& -
  2(\widehat{q}_1\cdot\widehat{q}_2)\frac{E}{m}+ \frac{2}{\vert
    q_1\vert \vert q_2\vert}
  v_1\cdot\Pi(q_1)\cdot\Pi(q_2)\cdot v_2
  -
  \nonumber
  \\
  &{}&
  \frac{\lambda}{m}(\widehat{q}_1\cdot\widehat{q}_2)\left[v_2\cdot\Pi(q_2)\cdot\widehat{q}_1
    - v_1\cdot\Pi(q_1)\cdot \widehat{q}_2\right].
\nonumber
\end{eqnarray}
We see that $\overline{\mathcal{S}\hbox{ol}}_{\mathbb{P}^{[n_F]}_L}$ is also
one-dimensional, and is also spanned by $\mathbf{P}_{(+)}$. 

\subsection{A Lagrangian with Local Conformal and
  Time-reparametization Invariance}

The action
\begin{equation}
  S_{3} := sm\int\left[s\left(\frac{d\widehat{q}}{dt}\right)^2\right]^{1/2} dt,
\nonumber\end{equation}
where $s=\pm1$, is invariant under both the local conformal
transformations, $q^a\to \alpha(\mathfrak{u}) q^a$, and the
reparametization of time $t\to \tau(t)$ where $\tau$ is a
monotonically increasing function of $t$. Then
\begin{equation}
  \mathbf{\Omega}_L = \frac{m}{\vert q\vert}
  \frac{P_{ab}(u)}{\sqrt{sv\cdot \Pi(q)\cdot
      v}}\mathbf{d}q^a\wedge\mathbf{d}v^b,
\nonumber\end{equation}
and $\mathbf{\Omega}_F=0$. Here, $a=1. \dots, D$,
\begin{equation}
  u_a=\frac{\Pi_{ab}(q)v^b}{\sqrt{sv\cdot\Pi(q)\cdot v}},
\nonumber\end{equation}
so that $u^2 = s$, while $P_{ab}(u) = \Pi_{ab}(q) -su_a u_b$. As such,
$\hbox{ker }\mathbf{\Omega}_L(\mathfrak{u}) = $ ker 
$\mathbf{\Omega}_M(\mathfrak{u})$. Both $\mathcal{C}$ and
$\mathcal{G}$ are two-dimensional, and are spanned by
\begin{equation}
  \mathbf{U}^q_{(1)} = 
  \widehat{q}\cdot\frac{\mathbf{\partial}\>\>\>}{\mathbf{\partial} q}, \quad
    \mathbf{U}^q_{(2)} =
  u\cdot\frac{\mathbf{\partial}\>\>\>}{\mathbf{\partial} q}, \quad
  \hbox{and}\quad 
    \mathbf{U}^v_{(1)} =
  \widehat{q}\cdot\frac{\mathbf{\partial}\>\>\>}{\mathbf{\partial} v}, \quad
    \mathbf{U}^v_{(2)} =
    u\cdot\frac{\mathbf{\partial}\>\>\>}{\mathbf{\partial} v},
\nonumber\end{equation}
respectively. It follows that $\hbox{dim }(\overline{\hbox{ker
  }\mathbf{\Omega}_L}/\mathcal{G})=2$. 

Because this system is fully constrained, $E=0$. As
$\mathbf{\Omega}_F=0$ as well, there are no
Lagrangian constraints. It follows that $\mathcal{S}\hbox{ym}\mathcal{L}$ is two
dimensional and spanned by $\mathbf{U}_{(1)}^q$ and
$\mathbf{U}_{(2)}^q$. As $\mathbf{\beta}=0$ as well, $\mathcal{S}\hbox{ym}$
is also two dimensional, and is also spanned by  
$\mathbf{U}^q_{(1)}$ and $\mathbf{U}^q_{(2)}$.   

We found in \cite{ADS2020} that $\overline{\mathbf{X}}_{L} = 0$. A
general SOELVF is then $\overline{\mathbf{X}}_{EL} = u^{1}(\mathfrak{u}) 
\left[\mathbf{U}^q_{(1)}\right] + u^{2}(\mathfrak{u})
\left[\mathbf{U}^q_{(2)}\right]$, with
$u^{n}(\mathfrak{u})\in\overline{\mathcal{F}}$ for $n=1,2$. It follows
that $\overline{\mathcal{S}\hbox{ol}}_{\mathbb{P}^{[n_F]}_L}$ 
is also two-dimensional, and is spanned by $\mathbf{U}^{q}_{(1)}$ and
$\mathbf{U}^q_{(2)}$ as well.  

\begin{table}
  \begin{tabular}{c|c|ccccc}
    \hline
    Action & Potential & $\>\>\>\overline{\hbox{ker }\mathbf{\Omega}_L}/\mathcal{G}\>\>\>$ &
    $\>\>\>\mathcal{S}\hbox{ym}\>\>\>$ & $\>\>\>\mathcal{S}\hbox{ym}\mathcal{L}\>\>\>$
    & $\>\>\> I_{\left[1\right]  }\>\>\>$ &
    $\>\>\>\overline{\mathcal{S}\hbox{ol}}_{\mathbb{P}^{[n_F]}_L} \>\>\>$ \\
    \hline
    {} & $V_{AS}(\hat{q}^a)$ & 1 & 1 & 1 & 0 & 1 \\
    $S_{1}$ & $V_{sph}(\vert q\vert)+V_{AS}(\hat{q}^a)$ & 1 & 1 & 0 & 1 & 0 \\
    {} & $V(q^a)$ & 1 & 0 & 0& 1& 0 \\
    \hline
    $S_{2}$ & $\frac{\lambda}{2}\left[\frac{q_1^a}{\vert q_2\vert}
  \frac{d\>\>}{dt}\left(\frac{q_{2a}}{\vert q_1\vert}\right) -
  \frac{q_2^a}{\vert q_1\vert}
  \frac{d\>\>}{dt}\left(\frac{q_{1a}}{\vert q_2\vert}\right)
  \right]$ & 2 & 1 & 1& 1& 1 \\
    \hline
    $S_{3}$ & $0$ & 2 & 2& 2& 0& 2\\
  \hline
  \end{tabular}
  \caption{A summary of the symmetries of the three examples
    considered in this paper. With the 
    exception of the $I_{[1]}$ column, the numerical entries are the
    dimensionality of the vector spaces listed along the first row. Notice
    the case where the Euler-Lagrange equations of motion has a
    generalized Lie symmetry while the action itself does not. In all
    three examples, $\hbox{dim
    }(\mathcal{S}\hbox{ym}\mathcal{L})=\hbox{dim
    }(\overline{\mathcal{S}\hbox{ol}}_{\mathbb{P}_L^{[n_f]}})$. 
  }
  \label{Table}
\end{table}

\section{Concluding Remarks\label{&Conc}}

That each generalized Lie symmetry of the action
contributes one arbitrary function to the SOELVF for a dynamical
system is known anecdotally, and is a result expected on physical
grounds. For almost regular Lagrangians, the
appearance in physics of a generalized Lie symmetry is due to a local
gauge symmetry of the 
dynamical system, and thus to the absence of a gauge\textemdash the
length of vectors for local conformal invariance, or a measure
for time for time-reparametization invariance\textemdash for some
dynamical property of the system. As the generalized Lie symmetries of
the action for an almost regular Lagrangian would have
$N_{\mathcal{S}\hbox{ym}\mathcal{L}}$ of these gauge freedoms, it is
reasonable that the absence of these gauges will result in an 
equal number of arbitrary functions in the SOELVF. An equal
number of terms to fix these gauges would then be needed to determine the
dynamics of the system uniquely. But while these expectations are
reasonable, up to now they have been fulfilled only on a case-by-case
basis. This is in great part because the analysis of dynamical systems
with a local gauge symmetry has traditionally been done using
constrained Hamiltonian mechanics. Such analysis relies on the
canonical Hamiltonian, however, and the connection
between the canonical Hamiltonian and the symmetries of the Lagrangian
is indirect at best, in contrast to the Lagrangian approach followed
here. Moreover, the process of determining the total Hamiltonian
for the system is often prescriptive, with results that are specific to
the system at hand. By focusing on the Lagrangian and on the Lagrangian
phase space, we have been able to show for all systems with an almost
regular Lagrangian that has a constant rank Lagrangian two-form,
a direct link between local gauge symmetries and its dynamics. In
particular, it establishes a link between the number of gauge
symmetries of the action and the number of arbitrary functions that
naturally appear in the evolution of such dynamical systems.  

As $\gamma_{\mathbf{P}}^{[1]}=0$ for any choice of
$\mathbf{P}\in\mathcal{S}\hbox{ym}\mathcal{L}$, the vectors in
$\mathcal{S}\hbox{ym}\mathcal{L}$ do not contribute to the
first-order constraint manifold $\mathbb{P}_L^{[1]}$, and as such do
not contribute to the Lagrangian constraint algorithm at this order, or at
any higher orders. It is for this reason that the
$N_{\mathcal{S}\hbox{ym}\mathcal{L}}$ arbitrary functions $u^m_{[1]}$
are not determined by the algorithm, and why these functions will
still contribute to $\overline{\mathbf{X}}_{EL}$ even after the algorithm has been
completed. It also means that if second- and higher-order Lagrangian constraints
are introduced, they are accidental and cannot be due to the local
gauge symmetries of the action. Interestingly, we have yet to find a dynamical system
with a Lagrangian that is both almost-regular and has a Lagrangian
two-form with constant rank where second- or higher-order Lagrangian
constraints are introduced. 

This impact of generalized Lie symmetries on the dynamics of particles
illustrates the inherent differences between the analysis of the
symmetries of regular Lagrangians and that of almost regular
Lagrangians. For regular Lagragians, the generator of the generalized
Lie symmetry (at times referred to as a global symmetry) gives rise to
a prolongation vector, and the action of this prolongation on the Lagrangian gives
the variation of the action, $\delta S$, under this symmetry. When the
Euler-Lagrange equations of motion are thenimposed, the conserved
quantity for this symmetry along the path given by the solution of these equations of
motion is then obtained. While the generator of the generalized 
Lie symmetry for the almost regular Lagrangian $\mathbf{g}_L$ does
give a prolongation vector \textbf{pr} $\mathbf{g}_L$ Eq.~$(\ref{prog})$, and
while the action of \textbf{pr} $\mathbf{g}_L$ on $L$ does give
$\delta S$, imposing the Euler-Lagrange equations of motion on $\delta
S$ in Eq.~$(\ref{Ae1})$ gives the vacuous statement $\delta
S=0$. Instead, the requirement that $\delta S=0$ for all paths on
$\mathbb{Q}$ gives the conditions that the  
generators of the symmetry must satisfy. This in turn shows that the
existence of these generators is due solely to the Lagrangian being
singular. These conditions then 
affect the dynamics of the system through
$\gamma_{\mathbf{P}}^{[1]}=0$, and in doing so, sets a lower bound to
the dimensionality of 
$\overline{\mathcal{S}\hbox{ol}}_{\mathbb{P}_L^{[n_f]}}$.  

We have found it quite difficult to construct more than one example of a
dynamical system that has an almost regular Lagrangian with both a
generalized Lie symmetry and a 
Lagrangian two-form with constant rank on $\mathbb{P}_L$. We have,
on the other hand, found it quite easy to construct examples of dynamical
systems that have an almost regular Lagrangian with a generalized Lie
symmetry and a Lagrangian two-form whose rank varies across
$\mathbb{P}_L$. Indeed, it is the latter case that is the more
prevalent one, and yet much of the results of this 
paper and a good portion of the results of our previous one \cite{ADS2020} relies
on the condition that the rank of the Lagrangian two-form be constant on
$\mathbb{P}_L$. This is even more concerning when we realize that these
more prevalent systems are expected, by their nature, to have much
richer dynamics and mathematical structures (indeed, we have found that
such systems often require the introduction of second- or higher-order
Lagrangian constraints), and yet it is not known
which of the results that have been shown to hold for systems with
constant rank Lagrangian two-forms will still hold when the rank varies
across $\mathbb{P}_L$. Determining the generalized Lie symmetries of
these systems; showing that the passage from the Lagrangian to the
Hamiltonian phase space is possible; and finding the links between
symmetry and dynamics is a necessity for future research.  

\begin{acknowledgments}
  This paper would not have been possible without the contributions by
  John Garrison, who provided much of the underlying symmetry analysis
  of the action used in \textbf{Section \ref{&A-Sym}}, and most
  of the essential mathematics in \textbf{Section
    \ref{&review}}. Publication made possible in 
  part by support from the Berkeley Research Impact Initiative (BRII)
  sponsored by the UC Berkeley Library. 
\end{acknowledgments}

\appendix*
\section{}

The Euler-Lagrangian equations of motion for the action $S_1$ is
\begin{equation}
0  =\frac{m}{\vert q\vert^3}\Pi_{ab}(q)\ddot{q}^b-\frac{2m}{\vert
    q\vert^3}(\widehat{q}\cdot\dot{q})
\Pi_{ab}(q)\dot{q}^b+\frac{\partial V}{\partial q^a}.
\label{a1}
\end{equation}
Contracting both sides of this equation with $\widehat{q}$ results in
the first-order Lagrangian constraint Eq.~$(\ref{g-con})$, and it is
clear that dynamics is only possible on this constraint
surface. Acting on Eq.~$(\ref{a1})$ with $\Pi_{ab}(q)$ gives
\begin{equation}
0  =\frac{m}{\vert q\vert^3}\Pi_{ab}(q)\ddot{q}^b-\frac{2m}{\vert
    q\vert^3}(\widehat{q}\cdot\dot{q})
\Pi_{ab}(q)\dot{q}^b+\Pi_a^b(q)\frac{\partial V}{\partial q^b},
\label{Ap}
\end{equation}
since $\Pi_{ac}(q)\Pi^c_b(q)=\Pi_{ab}(q)$. But in this case
$V(q^a)=V_{sph}(\vert q\vert) +V_{AS}(\hat{q})$, and as
\begin{equation}
  \Pi_a^b(q)\frac{\partial V_{Sph}}{\partial
    q^b}=\Pi_a^b(q)\frac{\partial \vert q\vert}{\partial
    q^b}V'_{Sph}(\vert q\vert) =0,
  \nonumber
\end{equation}
while the identity
\begin{equation}
  \frac{\partial\hat{q}^a}{\partial q^b}=\Pi^a_b(q),
\nonumber\end{equation}
ensures that
\begin{equation}
  \Pi_a^b(q)\frac{\partial V_{AS}}{\partial q^b}=\frac{\partial
    V_{AS}}{\partial q^a},
  \nonumber
\end{equation}
Eq.~$(\ref{Ap})$ thereby reduces to the same equations of motion for the
system as found for \textit{Case 1}. It is for this reason that
the two cases have same generalized Lie symmetry.


\begin{thebibliography}{60}
\bibitem{ADS2020}
    A. D. Speliotopoulos,
    Constrained dynamics: generalized Lie symmetries, singular
    Lagrangians, and the passage to Hamiltonian mechanics, 
    \textit{J.~Phys~Commun.},
    \textbf{4} 065002 (2020).

  \bibitem{Got1978}
    M. J. Gotay, J. M. Nester and G. Hinds,
    Presymplectic manifolds and the Dirac-Bergmann theory of constraints,
    \textit{Journal of Mathematical Physics},
    \textbf{19} 2388--2399 (1978) 10.1063/1.523597.

    
  \bibitem{Got1979}
     M. J. Gotay and J. M. Nester,
    Presymplectic lagrangian systems I: the constraint algorithm and the equivalence theorem,
    \textit{Annales de L'Institut Henri Poincare, Section A},
    \textbf{30}(2) 129--142 (1979)
    
    \bibitem{Got1980}
    M. J. Gotay and J. M. Nester,
    Presymplectic lagrangian systems II: the second-order problem,
    \textit{Annales de L'Institut Henri Poincare, Section A},
    \textbf{32}(1) 1--13 (1980).

   \bibitem{Car1990a}
    J. F. Cari{\~n}ena,
    Theory of singular Lagrangians, 
    \textit{Fortschritte der Physik},
    \textbf{38}(9) 641--679 (1990) 10.1002/prop.2190380902.

  \bibitem{Hen1992}
         M. Henneaux and C. Teitelboim,
    \textit{Quantization of Gauge Systems},
    (Princeton University Press, Princeton, New Jersey, 1992).

  \bibitem{Dir1950}
           P. A. M. Dirac,
           Generalized Hamiltonian dynamics,
    \textit{Canadian Journal of Mathematics},
    \textbf{2} 129--148 (1950) 10.4153/CJM-1950-012-1.

    \bibitem{Mun1989}
           M. C. Mu{\~n}oz-Lecanda,
           Hamiltonian systems with constraints: A geometric approach,
    \textit{International Journal of Theoretical Physics},
    \textbf{28}(11) 1405--1417 (1989) 10.1007/BF00671858.

    \bibitem{Lus2018}
           L. Lusanna,
           Dirac-Bergmann constraints in physics: Singular Lagrangians, Hamiltonian constraints and the Second Noether Theorem,
    \textit{International Journal of Geometric Methods in Modern Physics},
    \textbf{15}(10) 1830004 (2018), 10.1142/S0219887818300040.

    \bibitem{Pri1983}
        G. Prince,
        Toward a classification of dynamical symmetries in classical mechanics,
    \textit{Bulletin of the Australian Mathematical Society},
    \textbf{27} 53--71 (1983) 10.1017/S0004972700011485.

    \bibitem{Pri1985}
        G. Prince,
        A complete classification of dynamical symmetries in classical mechanics,
    \textit{Bulletin of the Australian Mathematical Society},
    \textbf{32} 299--308 (1985) 10.1017/S0004972700009977.

    \bibitem{Cra1983}
        M. Crampin,
        Tangent bundle geometry Lagrangian dynamics, 
    \textit{Journal of Physics A: Mathematical and General Physics},
    \textbf{16} 3755--3772 (1983) 10.1088/0305-4470/16/16/014.

    \bibitem{Car1991}
        J. F. Cari{\~n}ena, J. Fern{\'a}ndez-N{\'u}{\~n}ez and E. Mart{\'i}nez,
        A geometric approach to Noether's Second Theorem in time-dependent Lagrangian mechanics,  
    \textit{Letters in Mathematical Physics},
    \textbf{23} 51--63 (1991) 10.1007/BF01811294.

    \bibitem{Car1988b}
        J. F. Cari{\~n}ena and M. F. Ra{\~n}ada, 
        Noether's theorem for singular Lagrangians, 
    \textit{Letters on Mathematical Physics},
    \textbf{15} 305--311 (1988) 10.1007/BF00419588.

    \bibitem{Car1992}
        J. F. Cari{\~n}ena, E. Mart{\'i}nez and J. Fern{\'a}ndez-N{\'u}{\~n}ez,
        Noether's theorem in time-dependent Lagrangian mechanics, 
    \textit{Reports on Mathematical Physics},
    \textbf{31} 189--203 (1992) 10.1016/0034-4877(92)90014-R.

    \bibitem{Car2003}
        J. F. Cari{\~n}ena, J. Fern{\'a}ndez-N{\'u}{\~n}ez and M. F. Ra{\~n}ada,
        Singular Lagragians affine in velocities, 
    \textit{Journal of Physics A: Mathematical and General Physics},
    \textbf{36} 3789--3807 (2003) 10.1088/0305-4470/36/13/311.
     
    \bibitem{Car1993}
    J. F. Cari{\~n}ena and J. Fern{\'a}ndez-N{\'u}{\~n}ez,
    Geometric theory of time-dependent singular Lagrangians, 
    \textit{Fortschritte der Physik},
    \textbf{41}(6) 517--552 (1993).

        \bibitem{Car1990b}
      J. F. Cari{\~n}nena and E.~Martinez, in \textit{Summetries and Algebra Structures in Physics, Part 2: Integral Systems, Soli State Physics, and Theory of Phase Transitions}, edited by V. V. Dodonov and V. I. Man'ko, (Nova Science Publishers, New York, 1991) Chap.~Generalized Jacobi equation and inverse problem in classical mechanics, pp 84--98.

    \bibitem{Mar1992}
    G. Marmo, G. Mendella and W. M. Tulczyjew,
    Symmetries and constants of the motion for dynamics in implicit form,
    \textit{Annales de L'Institut Henri Poincare, Section A},
    \textbf{57}(2) 147--166 (1992).

    \bibitem{Gra2002}
    X. Gr{\'a}cia and J. M. Pons,
    Symmetries and infinitesimal symmetries of singular differential equations,
    \textit{Journal of Physics A: Mathematical and General Physics},
    \textbf{35} 5059--5077 (2002) 10.1088/0305-4470/35/24/306.

    \bibitem{Gra2005}
    X. Gr{\'a}cia and R. Mart{\'i}n,
    Geometric aspects of time-dependent singular differential equations,
    \textit{International Journal of Geometric Methods in Modern Physics},
    \textbf{2}(4) 597--618 (2005) 10.1142/S0219887805000697.

    \bibitem{Pop2017}
    L. Popescu,
    Symmetries of second order differential equations on Lie algebroids,
    \textit{Journal of Geometry and Physics},
    \textbf{117} 84--98 (2017) 10.1016/j.geomphys.2017.03.006.

  \bibitem{deL1995}
    M. de Le{\'o}n and D. M. de Diego,
    Symmetries and constants of the motion for singular Lagrangian systems,
    \textit{International Journal of Theoretical Physics},
    \textbf{35}(5) 975--1011 (1996) 10.1007/BF02302383.
 
    \bibitem{Dim2016}
    N. Dimakis, P. A. Terzis and T. Christodoulakis ,
    Contact symmetries of constrained quadratic Lagrangians,
    \textit{Journal of Physics: Conference Series},
    \textbf{670} 1--6 (2016) 10.1088/1742-6596/670/1/012021.

    \bibitem{Pop2009}
    M. Popescu,
    Totally singular Lagrangians and affine Hamiltonians,
    \textit{Balkan Journal of Geometry and Its Applications},
    \textbf{14}(1) 60--71 (2009)
             
    \bibitem{Pop2011}
    M. Popescu and P. Popescu,
    Totally singular Lagrangians and affine Hamiltonians of higher order,
    \textit{Balkan Journal of Geometry and Its Applications},
    \textbf{16}(2) 122--132 (2011).    
    
  \bibitem{Olv1993}
        P. J. Olver,
    \textit{Applications of Lie Groups to Differential Equations},
    (Springer-Verlag, New York, New York, 1993).

  \bibitem{Abr1978}
    R. Abraham and J. E. Marsden,
    \textit{Foundations of Mechanics, 2nd ed},
    (Addison-Wesley, Reading, Massachusetts, 1978).

\end{thebibliography}
\end{document}